\documentclass[aps,prx,twocolumn,amsmath,amssymb,superscriptaddress]{revtex4-2}

\usepackage{amsmath,amssymb, times}
\usepackage{mathtools}
\usepackage{color}
\usepackage{bm}
\usepackage{bbm}
\usepackage{graphicx}
\usepackage[normalem]{ulem}
\usepackage{enumitem}
\usepackage{makecell}
\usepackage{amsthm}
\usepackage{thmtools}

\usepackage{here}
\usepackage{amsthm}
\usepackage{thmtools}
\usepackage{blindtext}
\usepackage{array}
\usepackage{bbding}
\usepackage[colorlinks=true,citecolor=blue,linkcolor=blue]{hyperref}

\declaretheoremstyle[
shaded={bgcolor=\color{rgb}{0.9,0.9,0.9}}
]{definition}

\usepackage{tikz}
\definecolor{tensorcolor}{rgb}{1., 1., 1.}
\definecolor{btensorcolor}{rgb}{0.65,0.77,0.95}
\usetikzlibrary{matrix,positioning, decorations.pathreplacing,calligraphy,decorations.markings, knots, arrows.meta,decorations.markings}
\tikzset{
    mid arrow/.style args={#1}{
        decoration={
            markings,
            mark=at position #1 with {\arrow{Stealth}}
        },
        postaction={decorate}
    }
}

\begin{document}

\preprint{APS/123-QED}

\title{Chern Insulators as Fixed Points of Entanglement Renormalization}

\author{Hing Nin Chan}
\thanks{These authors contributed equally to this work.}
\affiliation{Department of Physics, Hong Kong University of Science and Technology, Clear Water Bay, Hong Kong SAR, China}
\affiliation{Center for Theoretical Condensed Matter Physics, Hong Kong University of Science and Technology, Clear Water Bay, Hong Kong SAR, China}
\author{Wenqing Xie}%
\thanks{These authors contributed equally to this work.}
\affiliation{Department of Physics, Hong Kong University of Science and Technology, Clear Water Bay, Hong Kong SAR, China}
\affiliation{Center for Theoretical Condensed Matter Physics, Hong Kong University of Science and Technology, Clear Water Bay, Hong Kong SAR, China}

\author{Sing Lam Wong}
\affiliation{Department of Physics and Astronomy, University of British Columbia, Vancouver, British Columbia V6T 1Z1, Canada}
\affiliation{Quantum Matter Institute, University of British Columbia, Vancouver, British Columbia V6T 1Z4, Canada}

\author{Hoi Chun Po}%
\email{hcpo@ust.hk}
\affiliation{Department of Physics, Hong Kong University of Science and Technology, Clear Water Bay, Hong Kong SAR, China}
\affiliation{Center for Theoretical Condensed Matter Physics, Hong Kong University of Science and Technology, Clear Water Bay, Hong Kong SAR, China}

\begin{abstract}

We develop a modified formulation of zipper entanglement
renormalization for gapped free-fermion systems
that enables exact state factorization. Unlike conventional entanglement renormalization,
zipper entanglement renormalization employs
disentangling unitaries, referred to as zipper unitaries,
with exponentially decaying spatial tails.
These unitaries disentangle exponentially localized
occupied and empty modes into a product-state sector
while preserving the nontrivial topology in the
remaining courier sector.
Applying zipper entanglement
renormalization to Chern insulators, we find numerically
that repeated renormalization iterations converge to nontrivial
fixed points with finite, nonzero correlation lengths,
while preserving the Chern number throughout the flow.
We further show that once a representative fixed-point is obtained, other states within the same gapped phase can be realized as fixed points through
suitably chosen zipper unitaries.
This fixed-point structure enables scale-invariant
state reconstruction, in which the same
momentum-dependent unitary transformation is reused
across RG scales through Brillouin-zone downsampling.
We demonstrate this construction for the Qi--Wu--Zhang
model on the square lattice and the Haldane model on
the honeycomb lattice. In both cases, the target
Chern-insulator ground states are reconstructed with
errors approaching machine precision, faithfully
reproducing their exponentially decaying real-space
correlations. 
However, this exponential decay of correlations
is highly sensitive to small perturbations in the
reconstruction and is not faithfully reproduced
by the approximate tensor-network representations
considered here, even with the layered structure of zipper entanglement
renormalization.

\end{abstract}

\maketitle

\section{\label{sec:intro} Introduction}
The renormalization group (RG)
\cite{Fisher1998Renormalization,Wilson1975renormalization} provides a
systematic framework for describing physics across length scales by
eliminating short-distance degrees of freedom while retaining
long-distance structure. In quantum many-body systems, however,
coarse-graining must also account for entanglement. The density-matrix
renormalization group (DMRG) \cite{White1992Density} addresses this by
retaining the dominant eigenstates of the reduced density matrix and is
remarkably successful for one-dimensional gapped systems
\cite{Verstraete2004Density,Verstraete_2008,Cirac2021Matrix,
Schollw_ck_2011}. 
To tackle critical systems, the multiscale entanglement renormalization ansatz (MERA) realizes entanglement-based coarse-graining through disentanglers, which extract short-range entanglement, followed by isometries which remove a subset of degrees of freedom by projection to a product state defined on the current RG scale
\cite{Vidal2007Entanglement,Vidal2008Class,
vidal2010entanglementrenormalizationintroduction,Pfeifer_2009,
Evenbly2010Entanglement,Evenbly2010Boundary,Corboz2010Simulation,
Silvi_2010,evenbly2013quantumcriticalitymultiscaleentanglement,
Evenbly2014Class,Evenbly2015Tensor}. 

Conventional MERA employs strictly local gates and
isometries. 
In dimensions higher than one, MERA
forms a subclass of projected entangled pair states
(PEPS) with a bond dimension independent of
system size \cite{Barthel_2010}.
Although PEPS and MERA-based circuits have been developed to approximate
Chern-insulator states~\cite{Wahl_2013, Niu_2024, wong2025entanglementrenormalizationcircuits2d}, no-go theorems constrain the exact representation
of gapped, translation-invariant chiral states
within these frameworks
\cite{Dubail_2015,Li_2019}.
These limitations suggest an obstruction to fully
disentangling a Chern-insulator state into a
product state through a single quasi-local
transformation. This motivates us to go beyond the PEPS framework
and apply a layered renormalization scheme
applicable even to gapped Chern-insulator states.
In this scheme, a subset of the degrees of freedom
is disentangled at each step, while the remaining
degrees of freedom are progressively renormalized
across successive layers.

For gapped free-fermion systems, localized
single-particle orbitals typically exhibit
exponentially decaying spatial tails, suggesting
that the natural disentangling transformations
should be exponentially quasi-local rather than
strictly finite-range. 
Motivated by this observation,
we apply zipper entanglement renormalization (ZER)
\cite{wong2022zipperentanglementrenormalizationfree}
to gapped free-fermion systems. ZER is a real-space
renormalization scheme that employs disentangling
unitaries with exponentially decaying spatial tails,
referred to as zipper unitaries.
The exponentially quasi-local nature of these
unitaries distinguishes ZER from existing
free-fermion renormalization approaches based on
finite-depth local circuits and wavelet constructions,
while providing a complementary framework
\cite{Evenbly2010Entanglement,Fishman2015Compression,
Evenbly2016Entanglement,Evenbly2018Representation,
Haegeman2018Rigorous}.

At each renormalization-group (RG) step, ZER disentangles selected frozen modes and leaves a reduced courier sector defining a lattice which is self-similar to the original. All the frozen and courier modes are exponentially localized on the current scale.

Originally introduced as an entanglement RG scheme with a controllable error~\cite{wong2022zipperentanglementrenormalizationfree}, here we show that ZER can be made numerically exact when the input state is gapped.
For Chern insulators, the nonzero Chern number imposes a topological obstruction to constructing a complete set of exponentially localized Wannier functions spanning the occupied subspace \cite{Brouder2007Exponential,Monaco_2018}. 
However, as ZER only disentangles a subset of the frozen fermion modes at any step, the Chern number remains nontrivial and is carried by the remaining courier sector~\cite{Cole2025Reduced}. 

We demonstrate ZER numerically for two representative Chern-insulator
models: the Qi--Wu--Zhang (QWZ) model \cite{Qi2006Topological} on the square
lattice and the Haldane model \cite{Haldane1988Model} on the honeycomb
lattice. In both cases, the target Chern-insulator ground states are
reconstructed to numerical precision, while their exponentially
decaying real-space correlations are accurately reproduced. 
Through incorporating an upsampling step, ZER can also be performed on an effectively infinite lattice. Interestingly, we show that repeated applications of the ZER procedure to gapped Chern insulator generically leads to a converging fixed-point state together with a converging set of frozen and courier modes. In other words, self-similarity emerges from the procedure, and the state flows to a fixed point which carries the original Chern number and displays exponentially decaying correlation functions. 
Our observations can be contrasted with conventional RG intuition, in which any fixed point typically carries either a zero or infinite correlation length.
We show that such fixed points with finite, non-zero correlation lengths become possible because of the quasi-local nature of the ZER unitary, which is consistent with earlier discussions concerning the construction of gapped chiral states by augmenting MERA with quasi-local evolution ~\cite{chu2023entanglementrenormalizationcircuitschiral}.

The emergence of a ZER fixed point naturally raises
the question of whether such a fixed point is
is special in any sense. 
To address this question, we show that a ZER fixed
point is defined relative to a particular choice
of zipper unitary, forming a fixed-point pair
consisting of the state and its associated
disentangling zipper unitary. Once a representative fixed-point pair is known, corresponding pairs for other free-fermion states within the same gapped phase
can, in principle, be constructed.
To demonstrate this, we explicitly construct
fixed-point zipper unitaries for the ground states
of the QWZ and Haldane models.

The scale-invariant structure of a ZER fixed-point pair then enables a state-reconstruction scheme: starting from a single-site
state, we can successively introduce ancilla product states and apply
the same inverse zipper unitary defined at each RG scale. 
We show that this procedure accurately reproduces the original target states in a way analogous to the construction of a critical state using scale-invariant MERA
\cite{Pfeifer_2009}, but here we capture a Chern insulator ground state thanks to the quasi-local nature of the zipper unitaries.

Although reversing the ZER procedure allows one to reconstruct
the original state to machine precision, in any realistic implementation on, say, a quantum simulator would necessarily invoke unavoidable errors.
Motivated by such concerns, we investigate the reconstruction performance when the zipper unitaries are only approximately realized. 
For both weak random onsite phase rotations
and tensor-network approximations of the zipper
unitaries, we find despite a relative high state fidelity can be achieved, the asymptotic decay of 
spatial correlations get modified from exponential to algebraic decay.
Such observations can be rationalized as a consequence of the fragility of the original exponential decay in an RG architecture, as a small error in the late RG time (long distance) leads to long-range errors which cannot be efficiently canceled by the short-distance mode transformations.

Our results highlight the potential of ZER as a framework for understanding and reconstructing topological free-fermion states across scales, while also revealing the sensitivity of their exponentially decaying correlations to small reconstruction errors.

\section{\label{sec:zer} Zipper entanglement renormalization}

We apply zipper entanglement renormalization (ZER)
\cite{wong2022zipperentanglementrenormalizationfree}
to gapped free-fermion systems on $d$-dimensional lattices. As in conventional entanglement renormalization,
each RG step combines $b^d$ unit cells into a single
unit cell of the coarse-grained lattice, producing
a sequence of lattices
$\{\mathcal{L}_0,\mathcal{L}_1,\mathcal{L}_2,\ldots\}$.
Consequently, each unit cell of $\mathcal{L}_{\tau}$
corresponds to a block of $b^{d\tau}$ unit cells
of the original lattice $\mathcal{L}_0$.
To maintain a fixed number of degrees of freedom
per unit cell across RG scales, a fraction $1/b^d$
of the single-particle modes is retained at each
step, while the remaining fraction $1-1/b^d$
is projected out.

The key distinction between ZER and conventional
entanglement renormalization lies in the structure
of the disentangling unitaries.
Whereas conventional approaches employ finite-depth
circuits composed of strictly local gates,
ZER permits quasi-local unitaries whose matrix
elements decay exponentially with spatial separation.
This additional flexibility allows exponentially
localized Wannier orbitals which are frozen in that 
they have definite occupations
(either fully occupied or empty).
As we will show, the frozen modes can be constructed 
to be completely disentangled from the
remaining degrees of freedom. This yields an exact
factorization of the state into a frozen
product-state sector and a courier sector
that carries the remaining entanglement.
Such an exact factorization is generally not
achievable using finite-depth circuits of strictly
local gates, since these circuits cannot completely
remove the exponentially decaying tails of
generic Wannier orbitals.

\subsection{Free-fermion systems}

The particle-number-conserving free-fermion systems are
described by quadratic Hamiltonians of the form
\[
H=\sum_{i,j}c_i^{\dagger}h_{ij}c_j,
\]
where $c_i^{\dagger}$ ($c_i$) creates (annihilates)
a fermion in the single-particle orbital $i$, and
$h_{ij}$ are the matrix elements of a Hermitian
single-particle Hamiltonian. In the absence of
interactions, the many-body ground state is a
fermionic Gaussian state, fully characterized by
the projector $P$ onto the occupied single-particle
subspace.

For a translation-invariant system, $P$ is block
diagonal in momentum space,
\[
P=\bigoplus_{\bm{k}\in\mathrm{BZ}}P(\bm{k}),
\]
where $P(\bm{k})$ is an $n\times n$ Hermitian
projector satisfying $P(\bm{k})^2=P(\bm{k})$.
We assume $n$ modes per unit cell, with
$n_o$ occupied and $n_e=n-n_o$ unoccupied modes,
so that $\mathrm{rank}\,P(\bm{k})=n_o$.

Under coarse-graining, the lattices
$\{\mathcal{L}_{\tau}\}$ have corresponding
Brillouin zones $\{\mathrm{BZ}_{\tau}\}$.
To describe the projectors at different
renormalization-group (RG) scales on a common
momentum domain, we introduce the dimensionless
crystal momentum
\begin{equation}
\tilde{\bm{k}}=a_{\tau}\bm{k}_{\tau}
\in\widetilde{\mathrm{BZ}}_{\tau},
\qquad
a_{\tau}=b^{\tau}a_0,
\end{equation}
where $a_{\tau}$ is the lattice spacing at
RG scale $\tau$.

In the thermodynamic limit,
all dimensionless Brillouin zones $\widetilde{\mathrm{BZ}} _{\tau}$ share a continuous domain. For a 2D square lattice, for instance, this domain is $[0,2\pi)^2$. For a finite
system with periodic boundary conditions, each
$\widetilde{\mathrm{BZ}}_{\tau}$ is a discrete
momentum grid within this domain. Under successive
coarse-graining, these grids become progressively
sparser, forming a nested sequence
\begin{equation}
\widetilde{\mathrm{BZ}}_{\tau+1}
\subseteq\widetilde{\mathrm{BZ}}_{\tau}
\subseteq\cdots\subseteq
\widetilde{\mathrm{BZ}}_0.
\end{equation}
This allows projectors at different RG scales
to be compared at common dimensionless momenta.

\subsection{Entanglement renormalization with quasi-local disentanglers}

A conventional entanglement-renormalization layer
consists of unitary disentanglers $u$ and
isometries, equivalently represented by coisometries
$w$ in the coarse-graining direction.
We adopt a formulation that separates the unitary transformations from
truncation by expressing each RG step as a global disentangling
unitary $U$ followed by a truncation map $R$.
In conventional MERA, the global disentangling
unitary $U$ is a depth-2
local circuit composed of inter-block disentanglers
$u$ and intra-block unitaries $v$.
This decomposition follows from the fact that
any $k\times n$ coisometry $w$, satisfying
$ww^\dagger=I_k$, can be written as
\begin{equation}
w=jv,\qquad
j=\begin{pmatrix}
I_k & \mathbf{0}_{k\times(n-k)}
\end{pmatrix},
\end{equation}
where $v$ is an $n\times n$ unitary and $j$ is the canonical truncation map that
retains the first $k$ components.

For truncation to preserve the purity of the
retained state, the discarded degrees of freedom
must be fully disentangled from the retained ones.
For particle-number-conserving free-fermion
Gaussian states, full disentanglement amounts
to separating fully occupied or empty
single-particle modes. The goal of the global
disentangling unitary $U$ is therefore to transform
such modes into strictly local orbitals within
each block, where they form an atomic-limit
product-state sector that can be removed by $R$.

For the atomic-insulator sector of a gapped
free-fermion state, the corresponding (fully
occupied or empty) Wannier
orbitals are exponentially localized. Mapping these orbitals onto strictly block-local  modes therefore generally
requires a unitary transformation with
exponentially decaying spatial tails. This
motivates extending the disentangler beyond
the finite-depth local circuits employed in
conventional MERA. We thus allow the global
disentangling unitary to be an exponentially
quasi-local unitary, whose single-particle
matrix elements decay exponentially with
distance, as discussed in
Appendix~\ref{app:exponentially_decaying}.

Despite their spatial tails, such transformations
remain sufficiently local to preserve gapped quantum
phases. Ground states connected by a uniformly gapped
Hamiltonian path are related by quasi-adiabatic
continuation
\cite{Chen2010Local,Hastings2005Quasiadiabatic}.
This continuation generates a quasi-local unitary
transformation whose spatial tails decay faster
than any power, although not necessarily
exponentially
\cite{Chen2010Local,Bravyi_2010}.
The exponentially quasi-local unitaries considered
in this work form a more restrictive class. In particular,
for translation-invariant free-fermion systems,
such unitaries correspond to analytic, periodic
transformations in momentum space and therefore
preserve the Chern number.

Motivated by these considerations, we employ
zipper entanglement renormalization (ZER)
\cite{wong2022zipperentanglementrenormalizationfree},
which allows disentangling transformations with
exponentially decaying spatial tails.
We refer to such transformations as zipper unitaries $Z$.
For the gapped free-fermion states considered here,
we construct zipper unitaries
that map localized occupied and empty modes
onto block-local product-state sectors.
As demonstrated below, this procedure enables exact
state factorization and reconstruction up to
machine precision.

\subsection{Procedures of zipper entanglement renormalization}\label{subsec:Procedures of zipper entanglement renormalization}

We now describe the zipper entanglement
renormalization (ZER) procedure in detail.
Compared with the original formulation in
Ref.~\cite{wong2022zipperentanglementrenormalizationfree},
we introduce a modified construction of the
zipper unitary $Z$ tailored to gapped
free-fermion systems. This construction ensures
that $Z$ is exponentially quasi-local and
exactly disentangles a subset of single-particle
modes, allowing their removal by the truncation
map.

At each RG step, the ZER procedure preserves
translation symmetry with respect to the
coarse-grained lattice. Consequently, different
crystal-momentum sectors remain decoupled,
allowing the disentangling transformation to
be constructed independently in each momentum
sector.

\subsubsection{Blocking}

At the RG step $\tau \rightarrow \tau+1$, we group $b^d$ unit cells
of $\mathcal{L}_{\tau}$ into a single block, which is identified as
one unit cell of the coarse-grained lattice $\mathcal{L}_{\tau+1}$.
Blocking does not change the physical state, but only reorganizes the
degrees of freedom into a new unit-cell basis. In momentum space, this
corresponds to folding the Brillouin zone: the $b^d$ momentum sectors
of $\widetilde{\mathrm{BZ}}_{\tau}$ are mapped onto the reduced Brillouin zone
$\widetilde{\mathrm{BZ}}_{\tau+1}$:
\begin{equation}
    \bigoplus_{\Tilde{\bm{k}} \in \Tilde{\mathrm{BZ}}_{\tau}}
    P_{\tau}(\Tilde{\bm{k}})
    \rightarrow
    \bigoplus_{\Tilde{\bm{k}} \in \Tilde{\mathrm{BZ}}_{\tau+1}}
    P_{\tau \rightarrow \tau+1}^{\mathrm{block}}
    (\Tilde{\bm{k}}).
\end{equation}
Since each original unit cell contains $n$ modes, each block contains
$nb^d$ modes. For each momentum in
$\mathrm{BZ}_{\tau+1}$, the $b^d$ momenta
$\tilde{\bm{k}}/b+\tilde{\bm{G}}_{\alpha}$, with $\alpha=1,\ldots,b^d$, in the original
Brillouin zone are folded onto the same crystal momentum $\tilde{\bm{k}}$ of the
coarse-grained lattice. Here, $\{\bm{G}_{\alpha}\}$ denotes the $b^d$
inequivalent reciprocal-lattice vectors of $\mathcal{L}_{\tau+1}$
modulo the reciprocal lattice of $\mathcal{L}_{\tau}$ and label the $b^d$ folded momentum sectors. In terms of the dimensionless momentum $\Tilde{\bm{k}}
    \in \Tilde{\mathrm{BZ}}_{\tau+1}$, the blocked
projector is given by
\begin{equation}
\begin{aligned}
    P_{\tau \rightarrow \tau+1}^{\mathrm{block}}
    (\Tilde{\bm{k}})
    &=
    \mathcal{F}^{-1}_{\mathrm{block}}
    \left[
    \bigoplus_{\alpha=1}^{b^d}
    P_{\tau}
    \left(
    \Tilde{\bm{k}}/b+\Tilde{\bm{G}}_{\alpha}
    \right)
    \right],
\end{aligned}
\end{equation}
where the Fourier transformation $\mathcal{F}_{\mathrm{block}}$ converts
between the folded momentum-sector basis and the basis of the $b^d$
unit cells within each block.

\subsubsection{Disentangling}

We construct a exponentially quasi-local  zipper unitary $Z_{\tau \rightarrow \tau+1}$ that
disentangles the blocked projector into a coarse-grained projector
$P_{\tau+1}$ and a product-state sector:
\begin{equation}
\label{eq: disentangle condition}
    Z_{\tau \rightarrow \tau+1}
    P_{\tau}^{\mathrm{block}}
    Z^{\dagger}_{\tau \rightarrow \tau+1}
    =
    P_{\tau+1} \oplus P_{\mathrm{prod}},
\end{equation}
where $P_{\mathrm{prod}}$ consists of fully occupied and empty localized modes
that are subsequently removed by the canonical removal map $R$. We assume that $Z_{\tau \rightarrow \tau+1}$ is translation invariant
with respect to the coarse-grained lattice $\mathcal{L}_{\tau+1}$.
It can therefore be decomposed into momentum sectors:
\[
  Z_{\tau \rightarrow \tau+1}
    =
    \bigoplus_{\Tilde{\bm{k}}
    \in \Tilde{\mathrm{BZ}}_{\tau+1}}
    Z_{\tau \rightarrow \tau+1}(\Tilde{\bm{k}}).
\]
Accordingly, the disentangling condition
Eq.~\eqref{eq: disentangle condition} becomes, in each momentum sector $\Tilde{\bm{k}}
    \in \widetilde{\mathrm{BZ}}_{\tau+1}$,
\begin{equation}
\begin{aligned}
    Z_{\tau \rightarrow \tau+1}(\Tilde{\bm{k}})
    P_{\tau}^{\mathrm{block}}(\Tilde{\bm{k}})
    Z^{\dagger}_{\tau \rightarrow \tau+1}(\Tilde{\bm{k}}) =
    P_{\tau+1}(\Tilde{\bm{k}})
    \oplus
    P_{\mathrm{prod}}(\Tilde{\bm{k}}),
\end{aligned}
\end{equation}
where
\begin{equation}
\label{eq: p_prod_k}
     P_{\mathrm{prod}}(\Tilde{\bm{k}})
    =
    \begin{pmatrix}
        0_{n_e(b^d-1) \times n_e(b^d-1)} & 0 \\
        0 & I_{n_o(b^d-1) \times n_o(b^d-1)}
    \end{pmatrix}.
\end{equation}

For critical states, the original ZER proposal constructs the frozen
modes by a two-step distillation procedure which extracts nearly decoupled
degrees of freedom by inspecting the entanglement spectrum of local reduced density matrices. For gapped states, we can further ensure numerically
exact decoupling by constructing the frozen mode Wannier functions though a projection of trial orbitals onto the occupied or empty subspaces. We note that a similar approach was discussed in Ref.~\cite{Cole2025Reduced} concerning how a subset of atomic degrees of freedom can be extracted from a topological band. 
Here, we extend this projection approach by constructing localized
orbitals for the occupied, empty, and remaining courier modes.

\begin{figure*}
    \centering
    \includegraphics[width=1\linewidth]{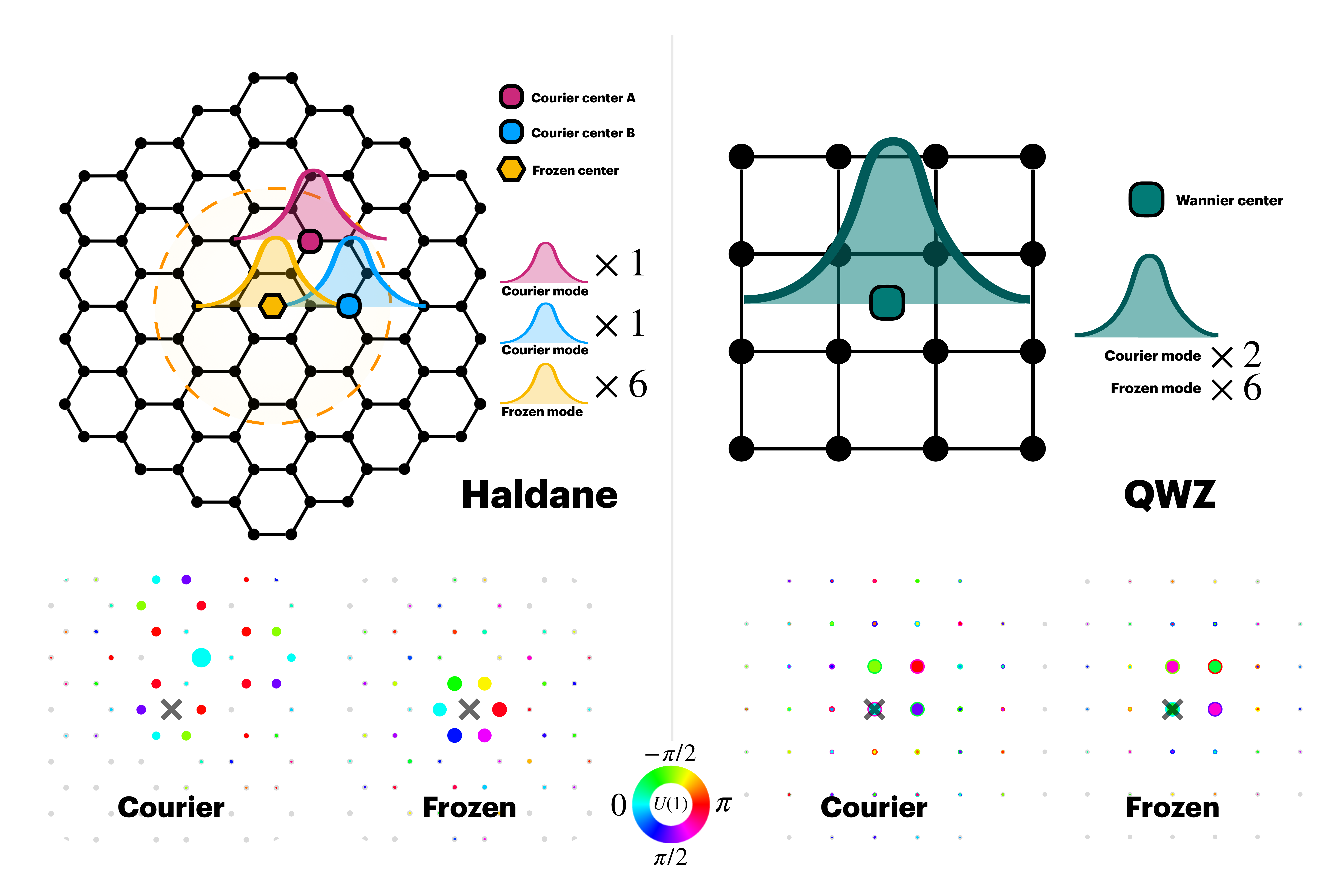}
    \caption{Symmetry-compatible choices of trial orbitals for the Haldane and QWZ models. We have plotted one of the courier and frozen Wannier functions from the perspective from the origin (indicated by the cross).
    (Left) For the Haldane model, the dashed orange circle indicates the local distilling region from which six finite-range trial orbitals for the frozen modes are selected, comprising three nearly occupied and three nearly empty modes. The two coarse-grained trial orbitals are centered at the two lattice sites of the honeycomb lattice at the next RG scale. This choice preserves the $C_3$ rotational symmetry and, for $\Delta=0$, is also compatible with the $C_6$ symmetry of the Haldane model.
    (Right) For the QWZ model, the trial orbitals are constructed as linear combinations of the modes within each block and are therefore centered at the block center. This choice preserves the $C_4$ rotational symmetry of the QWZ model.}
    \label{fig:honeycomb-qwz}
\end{figure*}

Each block contains $nb^d$ modes. We therefore choose $nb^d$ localized
trial orbitals within each block. For a Chern insulator, the
topological obstruction prevents the construction of a complete set of
exponentially localized Wannier functions spanning the occupied
subspace. Instead, we choose $n_o(b^d-1)$ trial orbitals
$|g_m^{\mathrm{occupied}}\rangle$ for the occupied modes and
$n_e(b^d-1)$ trial orbitals
$|g_m^{\mathrm{empty}}\rangle$ for the empty modes. The remaining
\[
    nb^d-(n_o+n_e)(b^d-1)=n
\]
trial orbitals $|g_m^{\mathrm{courier}}\rangle$ are assigned to the
courier sector. If the ZER procedure is required to preserve a certain symmetry, the trial orbitals must be chosen in a symmetry-compatible manner.  Examples of symmetry-compatible choices are illustrated in Fig.~\ref{fig:honeycomb-qwz}. Specifically, the trial orbitals for Haldane model preserve its $C_3$ rotational symmetry about the honeycomb sites, while those for the QWZ model preserve its $C_4$ rotational symmetry. These localized trial orbitals are then Fourier transformed into momentum space.

We project the occupied and empty trial orbitals onto their
respective target subspaces:
\begin{equation}
\begin{aligned}
    |\phi^{\mathrm{occupied}}_{m\bm{k}}\rangle
    &=
    P_{\tau}^{\mathrm{block}}(\bm{k})
    |g_{m\bm{k}}^{\mathrm{occupied}}\rangle, \\
    |\phi^{\mathrm{empty}}_{m\bm{k}}\rangle
    &=
    \left(I-P_{\tau}^{\mathrm{block}}(\bm{k})\right)
    |g_{m\bm{k}}^{\mathrm{empty}}\rangle.
\end{aligned}
\end{equation}
The projected orbitals are then orthonormalized independently within
the occupied and empty subspaces using L\"owdin orthonormalization
\cite{lowdin1950non}, yielding
$|\Tilde{\phi}^{\mathrm{occupied}}_{m\bm{k}}\rangle$ and
$|\Tilde{\phi}^{\mathrm{empty}}_{m\bm{k}}\rangle$. A brief review of
L\"owdin orthonormalization is provided in
Appendix~\ref{app:lowdin}.

The remaining $n$-dimensional subspace, which we refer to as the
courier subspace, is defined as the orthogonal complement of these
$n(b^d-1)$ occupied and empty modes:
\begin{equation}
\begin{aligned}
    P^{\mathrm{courier}}_{\tau}(\bm{k})
    = I
    &- \sum_{m=1}^{n_e(b^d-1)}
    |\Tilde{\phi}^{\mathrm{empty}}_{m\bm{k}}\rangle
    \langle \Tilde{\phi}^{\mathrm{empty}}_{m\bm{k}}| \\
    &- \sum_{m=1}^{n_o(b^d-1)}
    |\Tilde{\phi}^{\mathrm{occupied}}_{m\bm{k}}\rangle
    \langle \Tilde{\phi}^{\mathrm{occupied}}_{m\bm{k}}|.
\end{aligned}
\end{equation}
We then project the remaining $n$ trial orbitals onto the courier
subspace:
\begin{equation}
    |\phi^{\mathrm{courier}}_{m\bm{k}}\rangle
    =
    P^{\mathrm{courier}}_{\tau}(\bm{k})
    |g_{m\bm{k}}^{\mathrm{courier}}\rangle.
\end{equation}
Applying L\"owdin orthonormalization once more yields an orthonormal
basis
$|\Tilde{\phi}^{\mathrm{courier}}_{m\bm{k}}\rangle$
for the courier subspace.

Combining all $nb^d$ orthonormal orbitals, we construct the
zipper unitary as
\begin{equation}
    \label{eq:zipper_unitary}
    Z(\bm{k})
    =
    \begin{pmatrix}
        \Tilde{\Phi}^{\mathrm{courier}}_{\bm{k}} &
        \Tilde{\Phi}^{\mathrm{empty}}_{\bm{k}} &
        \Tilde{\Phi}^{\mathrm{occupied}}_{\bm{k}}
    \end{pmatrix}^{\dagger},
\end{equation}
where
\begin{equation}
    \Tilde{\Phi}^{\mathrm{courier}}_{\bm{k}}
    =
    \begin{pmatrix}
        \Tilde{\phi}^{\mathrm{courier}}_{1\bm{k}},
        \Tilde{\phi}^{\mathrm{courier}}_{2\bm{k}},
        \dots,
        \Tilde{\phi}^{\mathrm{courier}}_{n\bm{k}}
    \end{pmatrix}
\end{equation}
contains the courier orbitals as column vectors, while
$\Tilde{\Phi}^{\mathrm{empty}}_{\bm{k}}$ and
$\Tilde{\Phi}^{\mathrm{occupied}}_{\bm{k}}$ are defined analogously.

We next show that the zipper unitary constructed by the
projection and L\"owdin orthonormalization procedure is exponentially
localized. The key observation is that exponential localization in
real space is related to analyticity in momentum space \cite{des1964energy,Brouder2007Exponential}. 

For a gapped
free-fermion Hamiltonian with exponentially decaying real-space matrix
elements, the spectral projector onto the occupied bands is analytic
in $\bm{k}$ in a complex neighborhood of the Brillouin zone and
therefore decays exponentially in real space
\cite{des1964energy}. Since $P_{\tau}^{\mathrm{block}}(\bm{k})$ admits an analytic
continuation to a complex neighborhood of the Brillouin zone, and the
localized trial orbitals likewise have analytic Bloch representations,
their projections onto the occupied and empty subspaces remain analytic
in $\bm{k}$.

If the overlap matrices appearing in the L\"owdin orthonormalization (see Appendix~\ref{app:lowdin})
remain uniformly nonsingular,
\[
    \lambda_{\min}[S(\bm{k})] \ge \delta>0,
    \quad \text{for all } \bm{k},
\]
then $S^{-1/2}(\bm{k})$ is analytic in a possibly smaller complex
neighborhood. Hence, L\"owdin orthonormalization preserves analyticity.
The same argument applies to the construction and orthonormalization
of the courier subspace.

Consequently, all columns entering $Z(\bm{k})$ are analytic, and the
resulting zipper unitary is analytic in a complex neighborhood
of the Brillouin zone. Analyticity in momentum space then implies
exponential decay of its Fourier coefficients,
\begin{equation}
    \|Z(\bm{R})\|
    \le C e^{-\mu |\bm{R}|},
\end{equation}
for some $C,\mu>0$.

\subsubsection{Removing}

The zipper unitary $Z$ found above partitions the blocked state
into a courier sector containing $n$ modes and a product-state sector
containing $n(b^d-1)$ modes:
\[
    P_{\tau \rightarrow \tau+1}^{\mathrm{disent}}(\Tilde{\bm{k}})
    =
    P_{\tau+1}(\Tilde{\bm{k}})
    \oplus
    P_{\mathrm{prod}}(\Tilde{\bm{k}}),
\]
where $P_{\mathrm{prod}}$ consists of fully occupied and empty modes that are localized in each block, and
are disentangled from the retained courier sector.

To retain only the $n$ courier modes, we apply the canonical removal
map
\[
    R(\Tilde{\bm{k}})
    =
    \begin{pmatrix}
        I_n & \mathbf{0}_{n \times n(b^d-1)}
    \end{pmatrix},
\]
which selects the courier sector
\begin{equation}
    P_{\tau+1}(\Tilde{\bm{k}})
    =
    R(\Tilde{\bm{k}})
    P_{\tau \rightarrow \tau+1}^{\mathrm{disent}}
    (\Tilde{\bm{k}})
    R^{\dagger}(\Tilde{\bm{k}}),
\end{equation}
for all $\Tilde{\bm{k}} \in \Tilde{\mathrm{BZ}}_{\tau+1}$.

\subsubsection{Upsampling}

Since each RG step reduces the linear system size by a factor of $b$, a finite size lattice system reaches a single site after only a few RG steps unless one starts from a prohibitively large lattice. While this is sufficient for factorizing and reconstructing finite-size states, investigating
the behavior of the RG flow in the thermodynamic limit requires
additional iterations. To overcome this finite-size
limitation, we introduce an upsampling procedure
after each RG step, restoring the momentum-space
resolution of the coarse-grained projector to
that of the preceding RG scale:
\begin{equation}
    \bigoplus_{\tilde{\bm{k}} \in \tilde{\mathrm{BZ}}_{\tau+1}}
    P_{\tau+1}(\tilde{\bm{k}})
    \longrightarrow
    \bigoplus_{\tilde{\bm{k}} \in \tilde{\mathrm{BZ}}_{\tau}}
    P_{\tau+1}^{\mathrm{up}}(\tilde{\bm{k}}).
\end{equation}

The upsampling is performed by Wannier interpolation \cite{Marzari2012Maximally,Wang2006Abinitio}. We first Fourier
transform $P_{\tau+1}(\tilde{\bm{k}})$ to obtain its real-space
Fourier coefficients. These coefficients are centered and zero-padded
to enlarge the real-space lattice. An inverse Fourier transformation
then yields the interpolated projector
$P_{\tau+1}^{\mathrm{up}}(\tilde{\bm{k}})$ on the denser momentum
grid. Further details are given in
Appendix~\ref{app:wannier_interpolation}.

\section{\label{sec:state preparation}Every Chern insulator is a fixed point}

We apply the ZER procedure introduced above to
Chern-insulator models and observe that repeated
RG iterations converge toward a stationary state.
This motivates an operational definition of a
ZER fixed point in terms of the disentangling zipper
unitary $Z$. Remarkably, starting from a
Chern-insulating state, ZER can converge to a
nontrivial fixed point with a finite, nonzero
correlation length. This contrasts with the
conventional expectation that RG fixed points
have either vanishing or divergent correlation
lengths. As we discuss below, the existence
of such fixed points can be understood from
the quasi-local nature of the zipper unitary $Z$.

The ZER fixed point is, however, not uniquely
determined by the gapped phase. Although the zipper
unitary $Z_{\tau\rightarrow\tau+1}$ must be
exponentially quasi-local and fully disentangle
$n(b^d-1)$ modes from $P^{\mathrm{block}}_{\tau}$,
these requirements do not uniquely specify it.
This freedom allows for a broad class of ZER fixed
points: Any gapped Chern-insulator state connected
to a representative fixed point through a uniformly
gapped path of exponentially localized Hamiltonians
can also be realized as a ZER fixed point by
appropriately choosing a zipper unitary that
satisfies the same quasi-locality and disentangling
requirements.

\subsection{\label{subsec:Fix point}Fixed point of zipper entanglement renormalization}

\begin{figure*}
    \centering
    \includegraphics[width=1\linewidth]{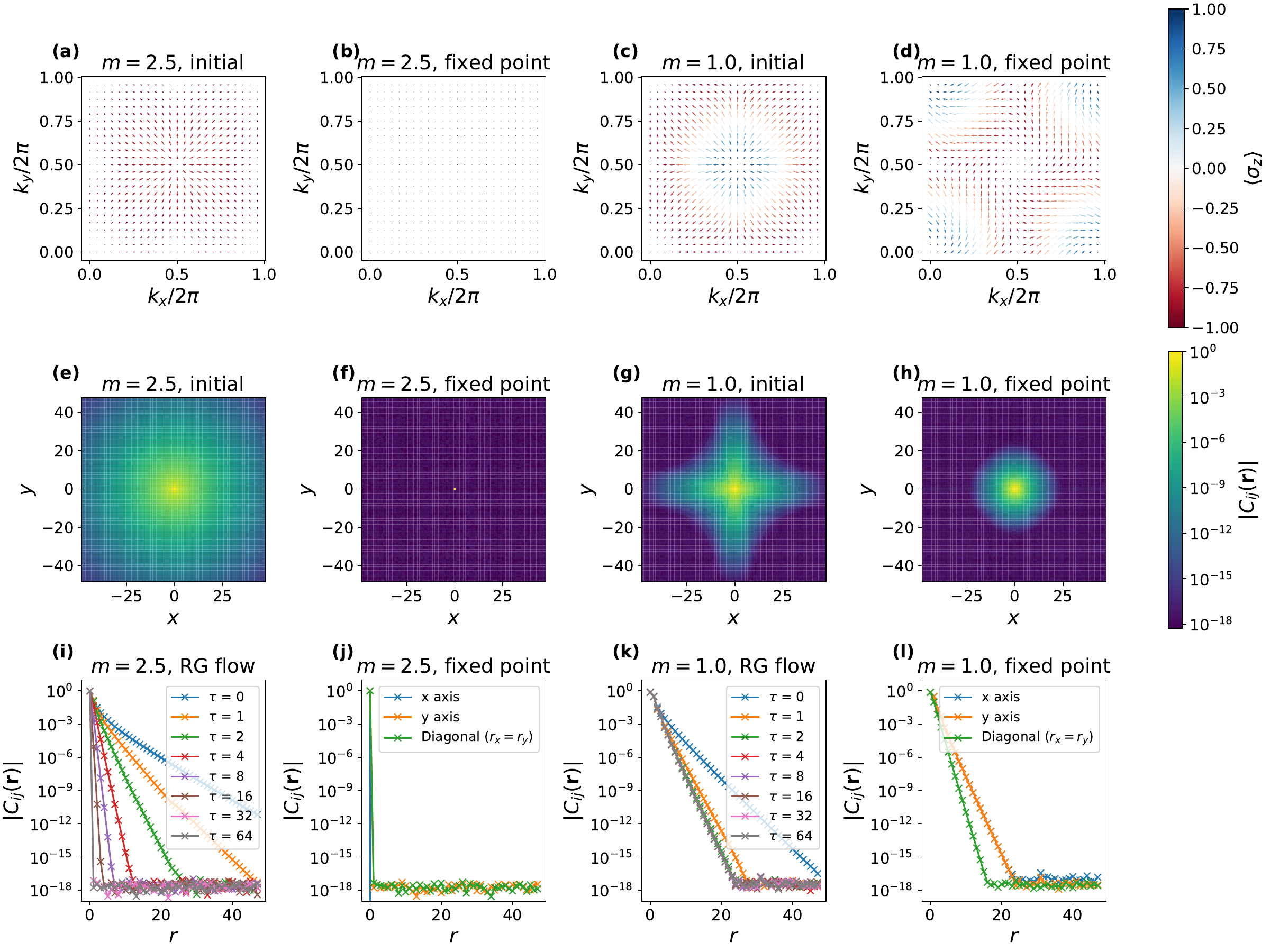}
    \caption{
        ZER flow and fixed points of the Qi--Wu--Zhang model.
        (a--d) Momentum-space pseudospin textures for the initial and
        fixed-point states at $m=2.5$ and $m=1.0$, respectively. The arrows
        represent the in-plane pseudospin components, while the color indicates
        $\langle \sigma_z \rangle$.
        (e--h) Real-space correlation magnitude $|C_{ij}(\bm{r})|$ for the
        corresponding initial and fixed-point states.
        (i,k) Evolution of $|C_{ij}(r)|$ under the ZER flow for $m=2.5$ and
        $m=1.0$, respectively, at different RG steps $\tau$.
        (j,l) Correlation functions of the corresponding fixed-point states
        along the $x$ axis, $y$ axis, and diagonal direction.
        For the trivial phase $m=2.5$, ZER flows to a product-state fixed point
        with zero correlation length, whereas for the Chern-insulating phase
        $m=1.0$, the fixed point retains a finite correlation length.
    }
    \label{fig: fix point}
\end{figure*}

We apply the ZER procedure to ground states of
the Qi--Wu--Zhang (QWZ) model~\cite{Qi2006Topological},
whose Bloch Hamiltonian is given by
\begin{equation}
\label{eq:QWZ hamiltonian}
H(\mathbf{k})
=
\sin k_x \, \sigma_x
+
\sin k_y \, \sigma_y
+
\bigl(m+\cos k_x+\cos k_y\bigr)\sigma_z .
\end{equation}
The band topology is controlled by the parameter $m$:
the system is in a Chern-insulating phase for $0<|m|<2$
and in a topologically trivial phase for $|m|>2$.
We consider two representative parameter values,
$m=1.0$ and $m=2.5$, corresponding to the Chern-insulating
and trivial phases, respectively.

Remarkably, with the upsampling procedure,
the ZER flow converges, with
the state $P_{\tau}$ approaching a fixed point
after several tens of RG steps. The Chern number
remains unchanged throughout the flow, and,
for a fixed ZER prescription, the corresponding
zipper unitary $Z_{\tau\rightarrow\tau+1}$
also converges.

As shown in Fig.~\ref{fig: fix point}, after 32 forward
RG steps on a $256\times256$ lattice, the trivial
state at $m=2.5$ flows to a product-state fixed point
with vanishing correlation length. In contrast, the
Chern-insulating state at $m=1.0$ flows to a
nontrivial fixed point with a finite, nonzero
correlation length. This behavior contrasts with
the conventional expectation that RG fixed points
have either vanishing or divergent correlation
lengths. As discussed below, the finite correlation
length is made possible by the quasi-local nature
of the zipper unitary. Similar fixed-point
behavior is also observed in the Haldane model
(see Sec.~\ref{subsec:Haldane}).

Even for a given initial state, the ZER fixed point
need not be unique. This nonuniqueness arises from
the freedom in constructing the zipper unitary
$Z_{\tau\rightarrow\tau+1}$. Although the
zipper unitary is required to be exponentially
quasi-local and to fully disentangle
$n(b^d-1)$ modes from $P^{\mathrm{block}}_{\tau}$,
these constraints do not uniquely determine it.
For example, different choices of trial orbitals
in Sec.~\ref{subsec:Procedures of zipper entanglement renormalization}
can yield different zipper unitaries and hence
different fixed points.

These observations motivate an operational definition
of a ZER fixed point in terms of the zipper unitary $Z$.
We define $P_{\tau}$ as a ZER fixed point of the zipper unitary
$Z_{\tau\rightarrow\tau+1}$ if the latter maps
$P_{\tau}$ to a direct sum of a Brillouin-zone-downsampled
copy $P_{\tau+1}$ and a product-state sector:
\begin{equation}
\label{eq: ZER fix point}
Z_{\tau\rightarrow\tau+1} P_{\tau}
Z_{\tau\rightarrow\tau+1}^{\dagger}
=
P_{\tau+1}\oplus P_{\mathrm{prod}}.
\end{equation}
Here, $P_{\tau+1}$ is obtained by Brillouin-zone
downsampling of $P_{\tau}$, namely,
\begin{equation}
P_{\tau+1}(\tilde{\bm{k}})
=
P_{\tau}(\tilde{\bm{k}}),
\qquad
\tilde{\bm{k}}\in\widetilde{\mathrm{BZ}}_{\tau+1}.
\end{equation}
The product-state sector $P_{\mathrm{prod}}$ consists
of fully occupied and empty modes removed by the
canonical removal map $R$.  Since the removal map
$R$ is fixed, a ZER fixed point is fully specified
by the pair $(Z_{\tau\rightarrow\tau+1},P_{\tau})$.

\subsection{Finite correlation length at the ZER fixed point}

Chern-insulator states with finite, nonzero correlation
lengths have also been identified as fixed points of
quasi-local renormalization schemes in
Ref.~\cite{chu2023entanglementrenormalizationcircuitschiral}.
The existence of such fixed points can be understood
from the quasi-local nature of the disentangling
transformation.

In conventional entanglement renormalization, an RG step
consists of a finite-depth local disentangling circuit
followed by coarse-graining. A finite-depth local circuit
can modify correlations only within a bounded spatial
range and cannot generate correlations at arbitrarily
large separations from a product state. In particular,
it cannot change the correlation length. Rescaling
the lattice by a factor $b$ reduces the correlation
length, measured in units of the renormalized lattice
spacing, according to
\begin{equation}
\label{eq: correlation length reduction}
\ell \rightarrow \ell/b.
\end{equation}
This scaling behavior is modified when the disentangling unitary
is quasi-local rather than strictly local.

In Ref.~\cite{chu2023entanglementrenormalizationcircuitschiral},
the quasi-local transformation is generated by
quasi-adiabatic evolution under a time-dependent
Hamiltonian. Unlike a finite-depth circuit of
strictly local gates, a quasi-local unitary has
spatial tails extending over arbitrarily large
distances, allowing it to modify long-distance
correlations. Such modifications can compensate
for the reduction of the correlation length under
coarse-graining, thereby allowing a fixed point
with a finite, nonzero correlation length.

The iterative ZER procedure provides a complementary
construction of such a quasi-local unitary. In
particular, the zipper unitary constructed above is guaranteed
to be exponentially quasi-local. As will be demonstrated
below, this construction is well suited to numerical
implementation and enables state reconstruction
with errors approaching machine precision.

\subsection{\label{subsec:A path to all other states}
From one fixed point to other states in the same phase}

The ZER fixed points are not restricted to particular
representative states. In fact, any gapped free-fermion
state in the same topological phase can be realized as
a ZER fixed point with a suitably chosen zipper
unitary. This follows from the covariance of the
fixed-point condition under exponentially quasi-local
unitary transformations. Given a fixed-point pair $(Z,P)$ satisfying
Eq.~\eqref{eq: ZER fix point}, we can construct another
fixed-point pair $(Z',P')$ through a unitary
transformation $S$:
\begin{equation}
\begin{aligned}
Z'_{\tau\rightarrow\tau+1}
&=
(S_{\tau+1}\oplus I)
Z_{\tau\rightarrow\tau+1}S_{\tau}^{\dagger},\\
P'_{\tau}
&=
S_{\tau}P_{\tau}S_{\tau}^{\dagger}.
\end{aligned}
\end{equation}
Here, identity $I$ acts on the product-state
sector, and $S_{\tau+1}$ is the corresponding
transformation on the downsampled sector.

To preserve the exponential quasi-locality of the
zipper unitary $Z'$, $S$ must also be exponentially
quasi-local. As we show below, such a unitary exists
for free-fermion states connected by a uniformly
gapped path of exponentially localized Hamiltonians.

Gapped states in the same phase can be connected
by quasi-adiabatic continuation
\cite{Chen2010Local,Hastings2005Quasiadiabatic}.
In general, the resulting unitary is quasi-local,
with spatial tails that decay faster than any
power but are not necessarily exponential
\cite{Chen2010Local,Bravyi_2010}. For free-fermion systems, a stronger locality
property can be established. Consider a uniformly
gapped path of Bloch Hamiltonians with exponentially
decaying real-space matrix elements. The associated
spectral projectors $P_s(\bm{R})$ also decay
exponentially in real space. The interpolation
between these projectors can be implemented by
Kato parallel transport~\cite{kato1950adiabatic}.
As shown in Appendix~\ref{app: Connecting},
the resulting single-particle unitary $S$ has
exponentially decaying real-space matrix elements.
Thus, free-fermion states connected by such a
gapped path admit an exponentially quasi-local
connecting unitary. Combining this unitary with
the original zipper unitary $Z$ yields a new
exponentially quasi-local zipper unitary $Z'$
for the transformed state $P'$.

In practice, Kato parallel transport can be
computationally cumbersome because it requires
evaluating a path-ordered exponential. A more
convenient alternative is the direct rotation
between the occupied subspaces described by
$P(\bm{k})$ and $P'(\bm{k})$
\cite{Bravyi_2011}:
\begin{equation}
S(\bm{k})
=
\left[
\bigl(I-2P'(\bm{k})\bigr)
\bigl(I-2P(\bm{k})\bigr)
\right]^{1/2},
\end{equation}
where the principal branch of the square root
is chosen. This construction gives the minimal
rotation between the two subspaces and is closely
related to geodesic constructions on the unitary
group~\cite{Hackl_2018}. It is well defined
provided that
\begin{equation}
\label{eq: overlap condition}
\|P'(\bm{k})-P(\bm{k})\|<1,
\qquad \forall\,\bm{k}.
\end{equation}
If this condition holds uniformly over the
Brillouin zone and both projectors are analytic
in momentum, then $S(\bm{k})$ is analytic in
a possibly narrower complex strip. Consequently,
its real-space matrix elements decay exponentially.
Further details are provided in
Appendix~\ref{app:geodesic_projector}.

When the overlap condition in
Eq.~\eqref{eq: overlap condition} is violated,
it may be useful to improve the overlap before
constructing the direct rotation. One possibility
is to apply a momentum-independent unitary
rotation $W$,
\[
P(\bm{k})
\rightarrow
WP(\bm{k})W^{\dagger},
\]
or to shift one of the projectors in momentum
space,
\[
P'(\bm{k})
\rightarrow
P'(\bm{k}+\Delta\bm{k}).
\]
The uniform rotation $W$ acts locally within
each unit cell, while the momentum shift is
implemented by the on-site unitary
$e^{i\hat{\bm{x}}\cdot\Delta\bm{k}}$.
Both operations preserve locality and can be
incorporated into the final connecting unitary
$S$. An appropriate combination of these
transformations may improve the overlap between
the two subspaces. If
Eq.~\eqref{eq: overlap condition} is then satisfied,
the direct rotation construction becomes applicable.

\subsection{But no state wears a crown}

In the present construction, the ZER fixed point is
not uniquely determined by the gapped phase but can
depend on both the initial state and the choice of
trial orbitals. As discussed above, any gapped
free-fermion state within a given phase can be
realized as a ZER fixed point with a suitably chosen
zipper unitary $Z$. This flexibility is
advantageous for state reconstruction, as it allows
a target state to be constructed from a product
state through a scale-invariant procedure, as will be 
discussed in Sec.~\ref{sec: State reconstruction}.

However, this flexibility also raises a conceptual
question about the role of fixed points in ZER.
If every state within a phase can serve as a ZER
fixed point, the fixed-point condition alone does
not select a distinguished representative of that
phase. In conventional RG, different states within
the same phase are expected to flow toward a common
fixed point as microscopic details become
irrelevant under coarse-graining. Such a fixed point
captures the universal properties of the phase
rather than the short-distance structure of a
particular state. This contrast motivates the
search for additional criteria to identify
distinguished fixed points within the ZER framework.
More discussion on this in
Sec.~\ref{sec: discussion}.

\section{State reconstruction}
\label{sec: State reconstruction}

Reversing the ZER procedure provides a layered
construction of a gapped free-fermion state.
Starting from a single-site state 
\[
P_T=P_0(\tilde{\bm{k}}=\bm{0}) = P_0(\bm{r}=\bm{0}),
\]
each reconstruction step introduces a product-state sector
$P_{\mathrm{prod}}$ and applies the inverse
zipper unitary $Z_{\tau\rightarrow\tau+1}^{\dagger}$,
eventually recovering the original state $P_0$.

By exploiting the properties of ZER fixed-point
pairs $(Z,P)$, this reconstruction can be made
scale invariant. We demonstrate this construction
for the Qi--Wu--Zhang and Haldane models, achieving
state-reconstruction fidelities approaching
machine precision.

\subsection{Scale-invariant reconstruction of fixed-point states}

A ZER fixed-point pair $(Z,P)$ provides a
scale-invariant reconstruction of the state $P$:
the same momentum-dependent zipper unitary $Z(\tilde{\bm{k}})$,
restricted to the appropriate Brillouin-zone grid,
is applied at every RG scale, producing the target free-fermion
state $P$ at progressively finer Brillouin zone grids, corresponding to larger system sizes.

To see this, consider a fixed-point state $P_0$
on the initial lattice $\mathcal{L}_0$.
In momentum space, the zipper unitary decomposes as
\[
Z_{0\rightarrow1}
=
\bigoplus_{\tilde{\bm{k}}\in\widetilde{\mathrm{BZ}}_1}
Z_{0\rightarrow1}(\tilde{\bm{k}}).
\]
The fixed-point condition in
Eq.~\eqref{eq: ZER fix point} then takes the form
\begin{equation}
\label{eq: disentangle in k space}
\mathcal{F}^{-1}_{\mathrm{block}}
\left[
\bigoplus_{\alpha=1}^{4}
P_0(\tilde{\bm{k}}/2+\tilde{\bm{G}}_{\alpha})
\right]
\overset{Z_{0\rightarrow1}(\tilde{\bm{k}})}
{\longrightarrow}
P_0(\tilde{\bm{k}})
\oplus P_{\mathrm{prod}}(\tilde{\bm{k}}),
\end{equation}
where we have used
$P_1(\tilde{\bm{k}})=P_0(\tilde{\bm{k}})$
for $\tilde{\bm{k}}\in\widetilde{\mathrm{BZ}}_1$.

On a finite lattice, successive Brillouin-zone
downsampling produces nested momentum grids,
$\widetilde{\mathrm{BZ}}_{\tau+1}
\subseteq\widetilde{\mathrm{BZ}}_{\tau}$.
The fixed-point condition is therefore preserved
at each RG scale by restricting the initial
momentum-dependent projector and zipper unitary:
\begin{equation}
\begin{aligned}
Z_{\tau\rightarrow\tau+1}(\tilde{\bm{k}})
&=Z_{0\rightarrow1}(\tilde{\bm{k}}),
\quad
\tilde{\bm{k}}\in\widetilde{\mathrm{BZ}}_{\tau+1},\\
P_{\tau}(\tilde{\bm{k}})
&=P_0(\tilde{\bm{k}}),
\quad
\tilde{\bm{k}}\in\widetilde{\mathrm{BZ}}_{\tau}.
\end{aligned}
\end{equation}
Repeated downsampling eventually reduces the
lattice to a single site at scale $T$, described by
$P_T=P_0(\tilde{\bm{k}}=\bm{0})$.
Reversing the RG steps then reconstructs $P_0$
by successively introducing product-state sectors
and applying the inverse disentanglers.

In the thermodynamic limit, the momentum grids
become continuous, and Eq.~\eqref{eq: disentangle in k space}
defines a functional fixed-point relation over
the dimensionless Brillouin zone, with the
RG-scale labels suppressed. The resulting
reconstruction is analogous to scale-invariant
MERA~\cite{Pfeifer_2009}, but employs exponentially
quasi-local disentangling zipper unitaries.

As shown in
Sec.~\ref{subsec:A path to all other states},
fixed-point pairs can be related by exponentially
quasi-local unitaries along uniformly gapped
paths of exponentially localized Hamiltonians.
Therefore, once a representative fixed-point pair
is known, scale-invariant reconstructions can,
in principle, be obtained for other gapped
free-fermion states within the same phase.

\subsection{\label{subsec:qwz} Qi-Wu-Zhang model}

\begin{figure*}
    \centering
    \includegraphics[width=1\linewidth]{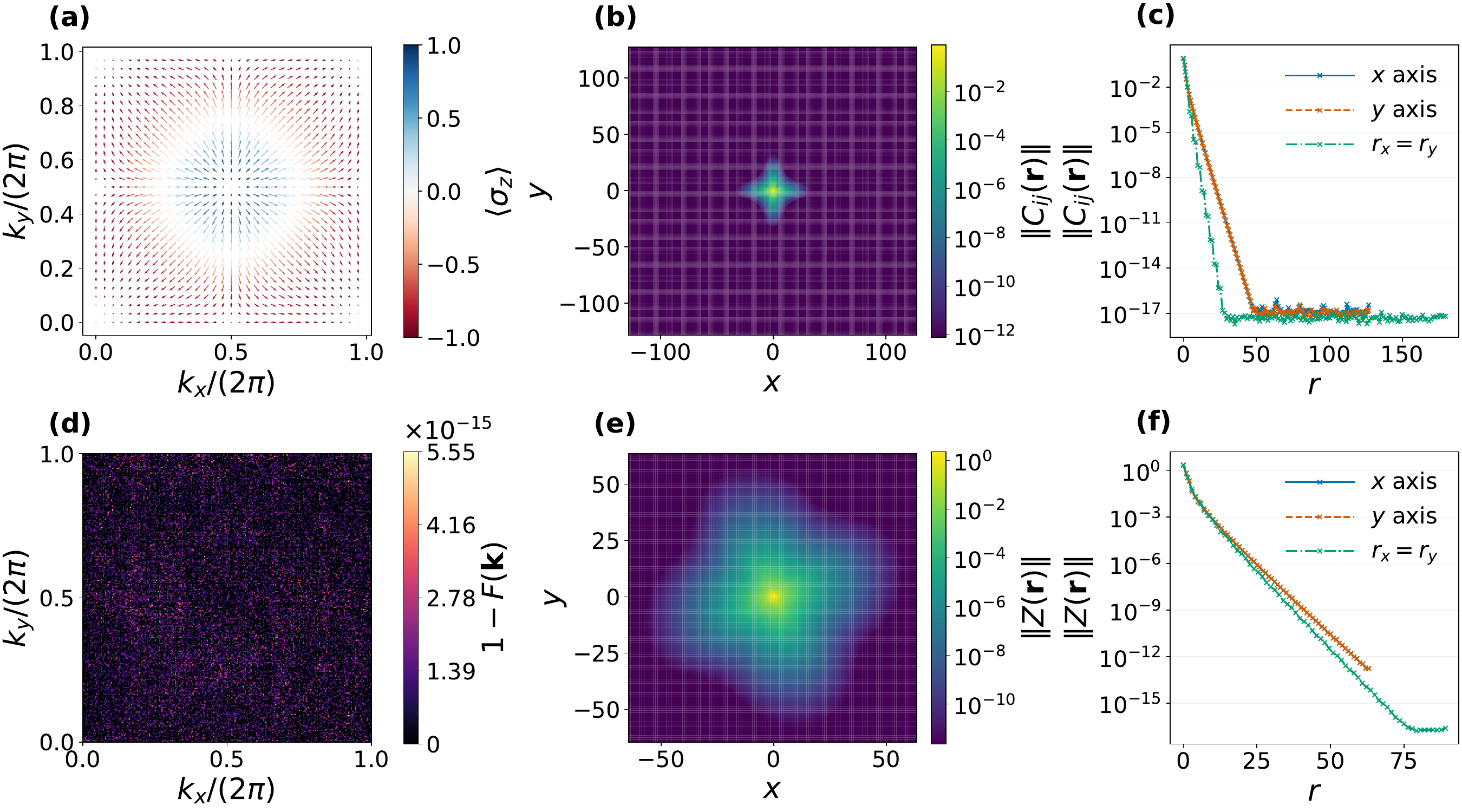}
    \caption{
    Reconstructed ground-state projector and spatial structure of the
    reconstruction unitary.
    (a) Pseudo-spin texture of the reconstructed projector in the Brillouin zone.
    Arrows show the in-plane components $\langle\sigma_x\rangle$ and
    $\langle\sigma_y\rangle$, while color indicates
    $\langle\sigma_z\rangle$.
    (b) Real-space correlations measured by the Frobenius norm
    $\|C_{ij}(\mathbf{r})\|_{\mathrm{F}}$.
    (c) Directional cuts of (b) along the $x$ axis, $y$ axis, and diagonal
    $r_x=r_y$, plotted against $r=|\mathbf{r}|$.
    (d) Momentum-resolved infidelity $1-F(\mathbf{k})$, where
    $F(\mathbf{k})=\mathrm{Tr}[P_{\mathrm{rec}}(\mathbf{k})
    P_{\mathrm{target}}(\mathbf{k})]$ compares the reconstructed and
    target ground-state projectors. Brighter colors indicate larger errors.
    (e) Real-space profile of the zipper unitary, quantified by
    $\|Z(\mathbf{r})\|_{\mathrm{F}}$.
    (f) Directional cuts of (e) along the same directions as in (c). The exponentially decaying tail
    of $C_{ij}(\mathbf{r})$ is accurately reconstructed. Moreover, the zipper unitary $Z(\mathbf{r})$ also exhibits exponential decay.
    }
    \label{fig:reconstruction}
\end{figure*}

As a demonstration, we apply the scale-invariant
reconstruction to the ground state $P^{\mathrm{tar}}$
of the Qi--Wu--Zhang model~\cite{Qi2006Topological}
(Eq.~\eqref{eq:QWZ hamiltonian}) at $m=1.0$,
which lies in the Chern-insulating.

We first iterate the ZER procedure to obtain a representative fixed-point pair $(Z^{*},P^{*})$ within the same phase, following the approach described in Sec.~\ref{subsec:Fix point}. The trial orbitals are chosen to be compatible with the $C_4$ rotational symmetry, ensuring that this symmetry is preserved throughout the RG flow; see Fig.~\ref{fig:honeycomb-qwz}. We then construct a unitary
$S$ connecting $P^{*}$ to $P^{\mathrm{tar}}$
using the direct-rotation method described in
Sec.~\ref{subsec:A path to all other states}.
However, the occupied subspaces of $P^{*}$
and $P^{\mathrm{tar}}$ are orthogonal at
$\tilde{\bm{k}}=(0,0)$ and $(\pi,\pi)$,
where the overlap condition for direct rotation
is violated. To circumvent this obstruction,
we first perform a momentum shift and then
apply the direct rotation. The corresponding
zipper unitary for the target state is
\begin{equation}
\label{eq: corresponding disentangling unitary}
Z^{\mathrm{tar}}
=
(S\oplus I)Z^{*}S^{\dagger},
\end{equation}
where $S$ includes the momentum shift and
subsequent direct rotation. The resulting
$Z^{\mathrm{tar}}$ exhibits exponentially
decaying real-space matrix elements, as shown
in Figs.~\ref{fig:reconstruction}(e) and
\ref{fig:reconstruction}(f).

With $Z^{\mathrm{tar}}$ determined, we reconstruct
$P^{\mathrm{tar}}$ from a single-site state
by successively introducing product-state
sectors $P_{\mathrm{prod}}$ and applying the
inverse zipper unitary
$Z^{\mathrm{tar}\dagger}$ at each RG scale.
As shown in Figs.~\ref{fig:reconstruction}(a)--(d),
the reconstructed state agrees with the target
state to near machine precision, with a maximum
momentum-resolved infidelity of approximately
$5.55\times10^{-15}$. The reconstruction also
faithfully reproduces the exponentially decaying
real-space correlations of the target state.

\subsection{\label{subsec:Haldane} Haldane model}

The same procedure can be applied to the Haldane model \cite{Haldane1988Model} on the
honeycomb lattice. In momentum space, the Bloch Hamiltonian can be
written as
\begin{equation}
    H(\bm{k})
    =
    d_0(\bm{k}) I
    +
    d_x(\bm{k}) \sigma_x
    +
    d_y(\bm{k}) \sigma_y
    +
    d_z(\bm{k}) \sigma_z ,
\end{equation}
with
\[
\begin{aligned}
    d_0(\bm{k})
    &=
    2t_2\cos\phi
    \sum_{j=1}^{3}\cos(\bm{k}\cdot\bm{b}_j), \\
    d_x(\bm{k})
    &=
    t_1\sum_{j=1}^{3}
    \cos(\bm{k}\cdot\bm{\delta}_j), \\
    d_y(\bm{k})
    &=
    -t_1\sum_{j=1}^{3}
    \sin(\bm{k}\cdot\bm{\delta}_j), \\
    d_z(\bm{k})
    &=
    \Delta
    -
    2t_2\sin\phi
    \sum_{j=1}^{3}\sin(\bm{k}\cdot\bm{b}_j),
\end{aligned}
\]
where $\bm{\delta}_j$ and $\bm{b}_j$ denote the three
nearest-neighbor and next-nearest-neighbor vectors, respectively,
$t_1$ and $t_2$ are the corresponding hopping amplitudes, $\phi$ is
the phase of the complex next-nearest-neighbor hopping, and $\Delta$
is the sublattice-staggered potential.

The general Haldane Hamiltonian preserves $C_3$ rotational symmetry.
When $\Delta=0$, the sublattice symmetry is restored and the
Hamiltonian further preserves the $C_6$ rotational symmetry of the
honeycomb lattice. To preserve the $C_3$ (or $C_6$) rotational symmetry of the Haldane
ground state under the ZER procedure, the trial orbitals must be chosen
in a symmetry-compatible manner. We choose the six frozen trial
orbitals, consisting of three occupied and three empty orbitals, from
the hexagonal distilling region shown in
Fig.~\ref{fig:honeycomb-qwz}. The remaining two courier
trial orbitals are chosen at the two lattice sites of the coarse-grained lattice at the next RG scale. 

To determine the occupied and empty trial orbitals within the
hexagonal region, we restrict the occupied-state projector to this
region, obtaining the restricted projector $P^{r}$. Its eigenvalues
$\epsilon_\alpha$ quantify the occupation weight of the corresponding
localized eigenmodes. Eigenvalues close to $1$ identify modes that are
predominantly occupied, whereas eigenvalues close to $0$ identify modes
that are predominantly empty. We therefore choose the three
eigenstates with the largest eigenvalues as the occupied trial
orbitals and the three eigenstates with the smallest eigenvalues as
the empty trial orbitals, as illustrated in
Fig.~\ref{fig:honeycomb-qwz}.

\begin{figure*}
    \centering
    \includegraphics[width=1\linewidth]{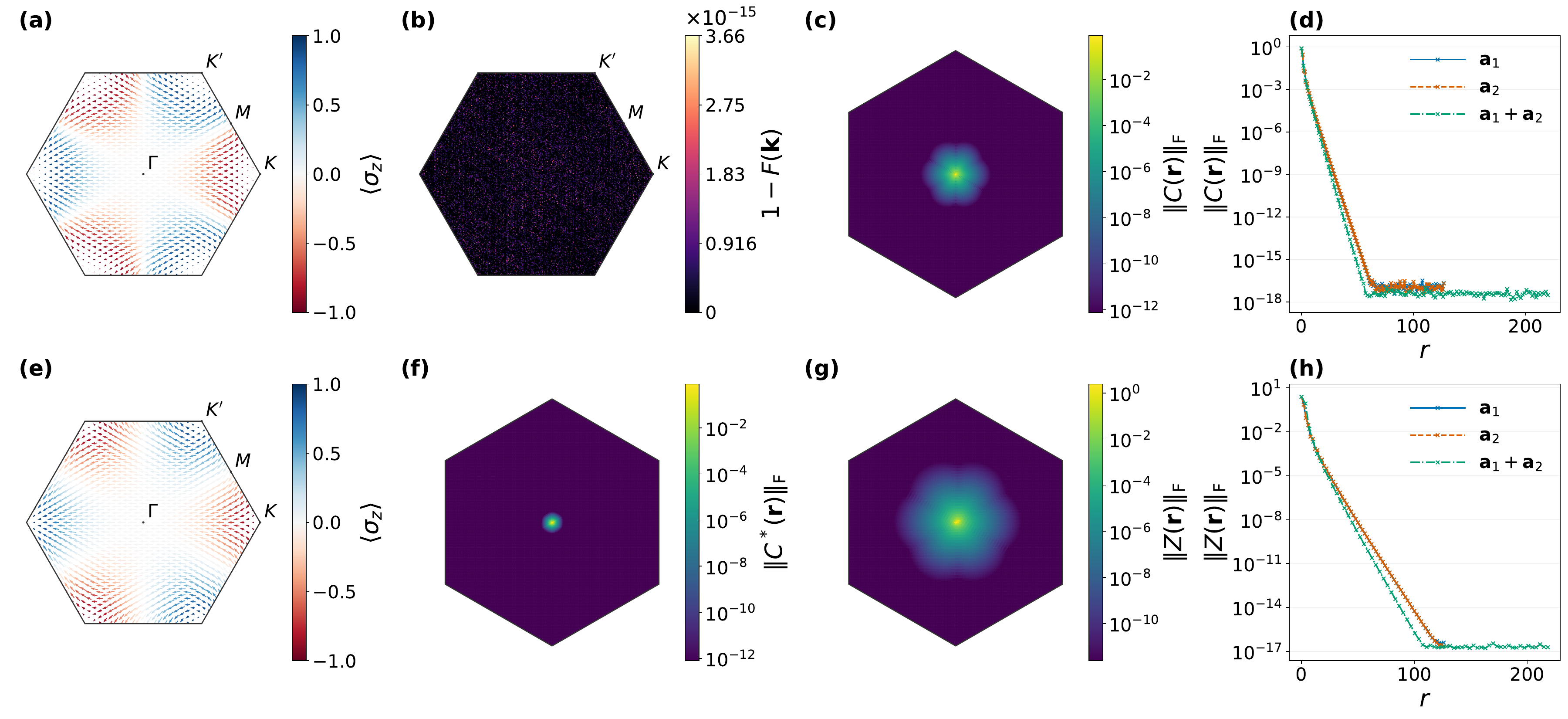}
    \caption{
Reconstruction of the Haldane-model ground state for
$t_1=1$, $t_2=0.5$, $\phi=\pi/2$, and $\Delta=0$ on a $2^8\times2^8$ lattice with $8$ inverse ZER layers.
(a) Reconstructed pseudospin texture in the first Brillouin zone:
arrows show $(\langle\sigma_x\rangle,\langle\sigma_y\rangle)$,
and color denotes $\langle\sigma_z\rangle$.
(b) Momentum-resolved infidelity,
$1-F(\mathbf{k})=1-\operatorname{Tr}
[P_{\mathrm{rec}}(\mathbf{k})P_{\mathrm{target}}(\mathbf{k})]$,
with a maximum of approximately $3.55\times10^{-15}$.
(c,d) Real-space reconstructed correlations and their directional decay.
The correlations exhibit exponential decay in real space, as expected
for a gapped Chern insulator.
(e,f) Pseudospin texture and real-space correlations of the fixed point state
obtained after 32 forward RG steps.
(g,h) Zipper unitary in real space $Z(\mathbf{r})$
and its directional decay.
Spatial maps and directional line cuts display Frobenius norms,
with logarithmic color scales for the maps and logarithmic vertical
axes for the cuts. The cuts along $\mathbf{a}_1$, $\mathbf{a}_2$,
and $\mathbf{a}_1+\mathbf{a}_2$ reveal exponential decay down to
numerical precision.
}
    \label{fig:haldane_result}
\end{figure*}

We perform the ZER procedure and scale invariant state reconstruction for Haldane model at
\[
    t_1=1.0,\qquad
    t_2=0.5,\qquad
    \phi=\pi/2,\qquad
    \Delta=0.
\]
These parameters realize a Chern-insulating phase with Chern number $C=1$ and $C_6$ rotational symmetry.

The numerical results are shown in Fig.~\ref{fig:haldane_result}. The
Haldane-model ground state is reconstructed to numerical precision using eight inverse RG layers on a $2^8\times2^8$ lattice. The reconstructed occupied-band projector agrees with the target throughout the Brillouin zone, with a maximum infidelity of approximately $3.77\times10^{-15}$. Both the reconstructed correlation matrix and the zipper unitary $Z$ exhibit exponential decay in their real-space Frobenius norms. These results demonstrate the accuracy of the reconstruction.

\section{Approximation in the state reconstruction}

Thus far, state reconstruction has been demonstrated
to be exact up to machine precision. In practical implementations, however,
such as state preparation on quantum devices,
imperfections inevitably introduce deviations
from the ideal zipper unitaries. This
motivates us to investigate the robustness of
the scale-invariant reconstruction against
unitary perturbations.

As we show below, even when the reconstructed
state retains a high fidelity with the target
state, small perturbations can qualitatively
alter its long-distance correlations. In
particular, the exponential decay of correlations
in the target state is replaced by algebraic
decay in the presence of the noise considered
here. This sensitivity suggests that accurately
reproducing exponentially decaying correlations
may pose a significant challenge for numerical
optimization of tensor-network ansatzes. 
\cite{Dubail_2015}, even when employing the
layered structure of ZER.

\subsection{Adding noise}

\begin{figure}
    \centering
    \includegraphics[width=0.8\linewidth]{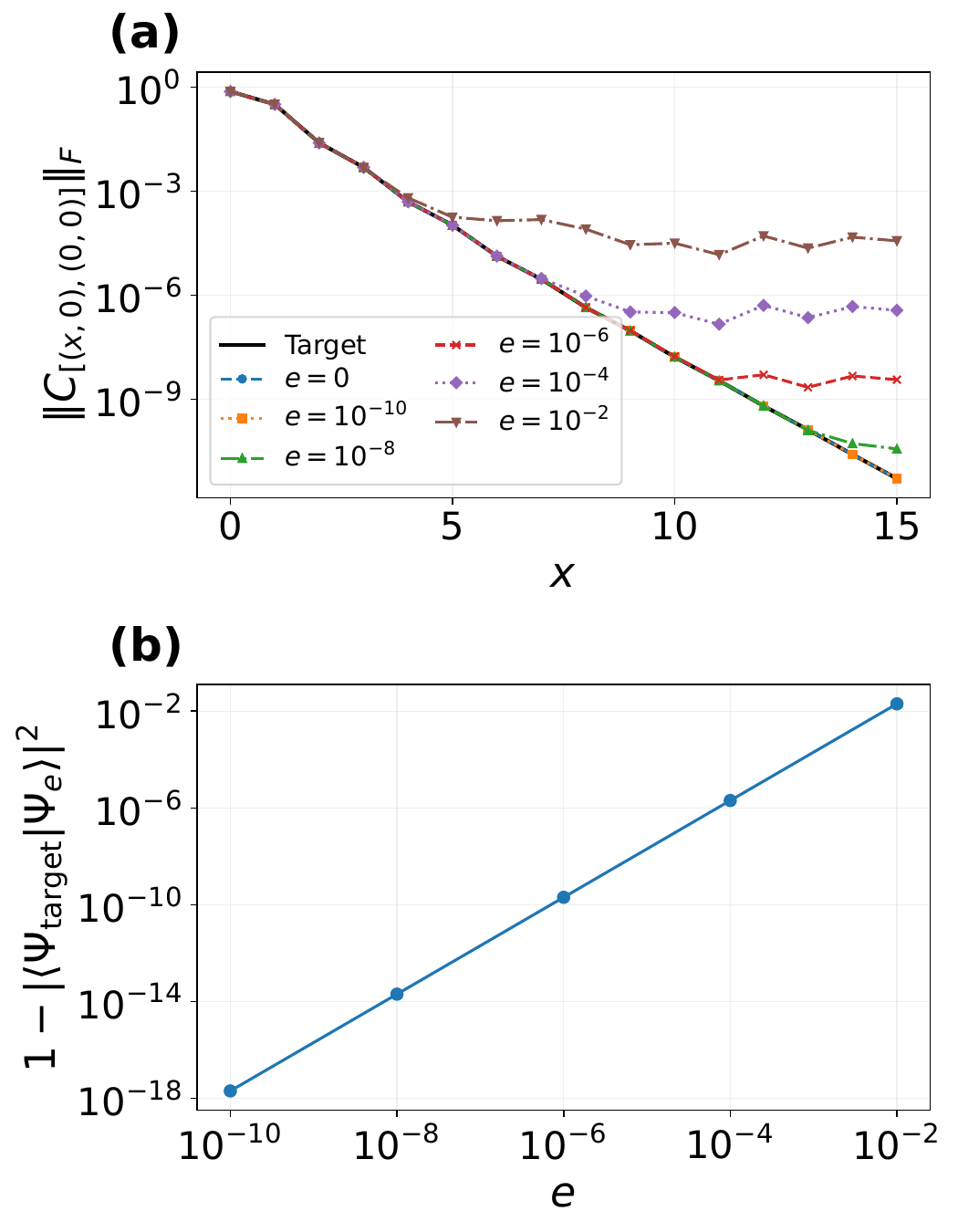}
    \caption{
    Effect of random on-site phase perturbations on
    scale-invariant state reconstruction.
    After each of the five inverse RG layers, we apply
    independent orbital-dependent phase perturbations
    $V_{\tau}=\bigoplus_{x,y,a}
    e^{i\epsilon\xi_{\tau,xya}}$,
    where $\xi_{\tau,xya}$ are independent random
    variables uniformly distributed in $[-1,1]$,
    and $\epsilon$ controls the perturbation strength.
    The same random realizations are used for all
    values of $\epsilon$.
    (a) Real-space correlation function
    $\|C_{[(x,0),(0,0)]}\|_F$ along the $x$ direction,
    compared with that of the target state.
    Increasing $\epsilon$ produces a long-distance
    correlation floor, disrupting the exponential
    decay of the target correlations.
    (b) Many-body infidelity
    $1-|\langle\Psi_{\mathrm{target}}|
    \Psi_{\epsilon}\rangle|^2$
    as a function of $\epsilon$.
    }
    \label{fig:add noise}
\end{figure}

To investigate the robustness of state reconstruction,
we introduce random on-site phase perturbations
at each inverse RG layers. Specifically,
we apply diagonal unitaries
\begin{equation}
V_{\tau}
=
\bigoplus_{x,y,a}
e^{i\epsilon\xi_{\tau,xya}},
\end{equation}
where $\xi_{\tau,xya}$ are independent random
variables uniformly distributed in $[-1,1]$,
and $\epsilon$ controls the perturbation strength.
Since these perturbations act locally on individual
orbitals, they cannot by themselves generate
correlations between spatially separated sites.

As shown in Fig.~\ref{fig:add noise}, despite
the high fidelity of the reconstructed state,
the exponentially decaying correlations of the
target state are disrupted, giving rise to
long-distance tails consistent with algebraic
decay. This demonstrates that high reconstruction
fidelity does not necessarily guarantee the
correct long-distance correlation behavior.

This sensitivity can be understood from the
multiscale nature of the reconstruction.
Correlations introduced at early reconstruction
layers correspond to large spatial separations
when measured on the original lattice.
To recover a state with a finite correlation
length, contributions from different RG scales
must combine to suppress these long-distance
correlations. In the ideal reconstruction,
this cancellation is achieved by the precisely
constructed inverse zipper unitaries. Small
perturbations disrupt the cancellation, leaving
residual long-distance correlations that can
survive through subsequent reconstruction steps.
Consequently, even strictly local perturbations
at individual RG layers can produce substantial
changes in the long-distance correlation behavior
of the reconstructed state.

\subsection{Tensor network ansatz for ZER}

\begin{figure*}
    \centering
    \includegraphics[width=1.0\linewidth]{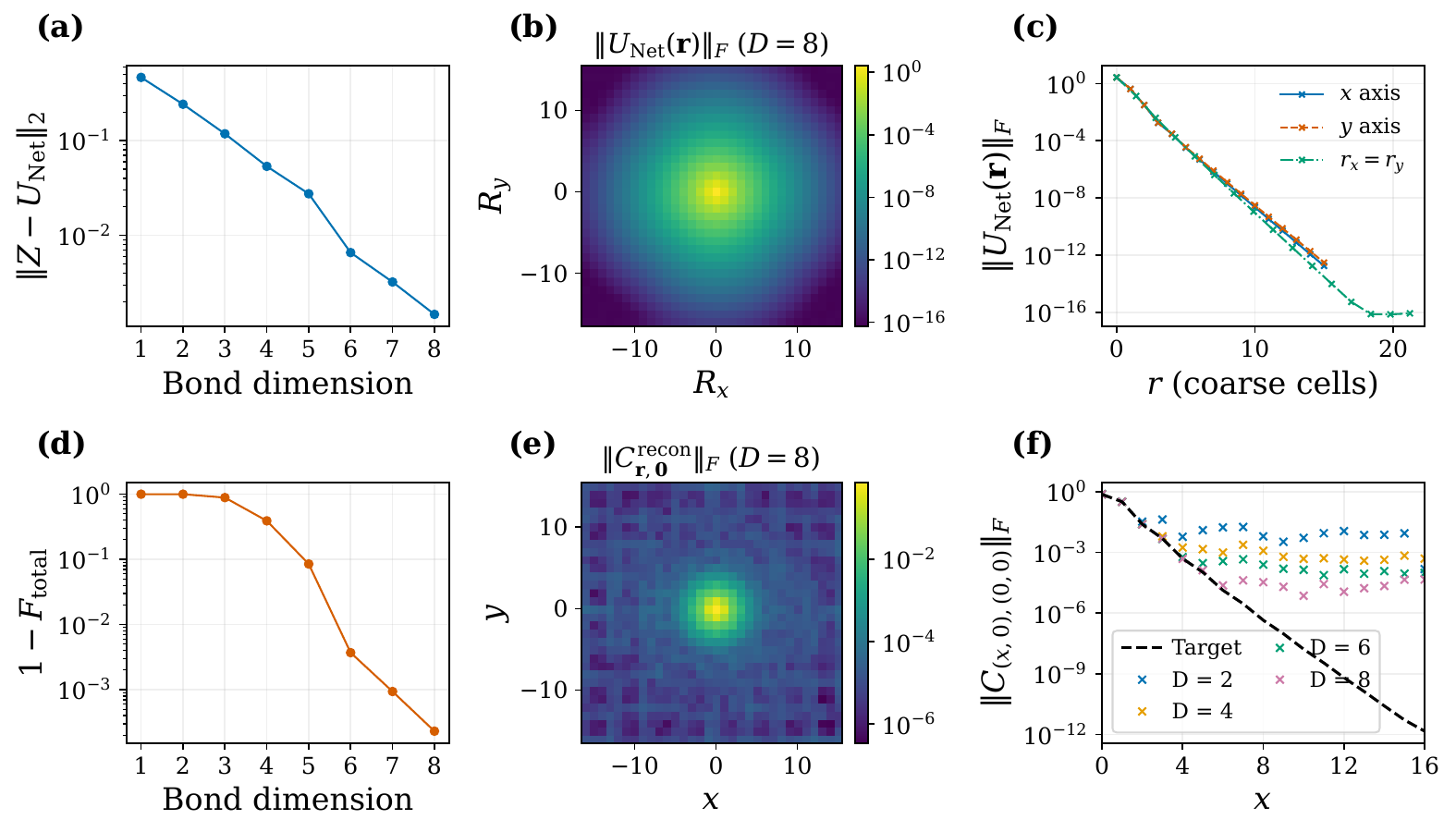}
   \caption{
Reconstruction of the Chern insulator state using a unitary network.
\textbf{(a)} Operator-norm difference
$\|Z-U_{\mathrm{Net}}\|_2$ between the target zipper unitary $Z$
and the unitary network $U_{\mathrm{Net}}$, as a function of bond dimension.
\textbf{(b)} Real-space kernel of the unitary network with bond dimension $D=8$, exhibiting an exponentially decaying tail in its Frobenius norm
$\|U_{\mathrm{Net}}(\mathbf{R})\|_F$.
\textbf{(c)} Cuts of (b) along the $x$ axis, $y$ axis, and
diagonal, with distance measured in coarse-lattice units.
\textbf{(d)} Total state infidelity $1-F_{\mathrm{tot}}$,
where
$F_{\mathrm{tot}}
=|\langle\Psi_{\mathrm{target}}|\Psi_{\mathrm{recon}}\rangle|^2$.
\textbf{(e)} Correlations $\|C^{\mathrm{recon}}_{\mathbf{r},\mathbf{0}}\|_F$ of the reconstructed state at bond
dimension $D=8$.
\textbf{(f)} Correlations along the $x$ direction for bond
dimensions $2,4,6,8$, compared with the target
(black dashed line).
}
\label{fig:un-reconstruction}
    \label{fig:un_reconstruction}
\end{figure*}

Tensor-network ansatzes provide systematic
approximations to quantum states and operators,
with their expressive power controlled by the
bond dimension. In particular, entanglement
renormalization gives rise to tensor-network
architectures such as the multiscale entanglement
renormalization ansatz (MERA)
~\cite{Vidal2007Entanglement,Vidal2008Class,
vidal2010entanglementrenormalizationintroduction,Pfeifer_2009,
Evenbly2010Entanglement,Evenbly2010Boundary,Corboz2010Simulation,
Silvi_2010,evenbly2013quantumcriticalitymultiscaleentanglement,
Evenbly2014Class,Evenbly2015Tensor}.
In ZER, the disentangling unitary is generalized
from a finite-depth local unitary circuit to an
exponentially quasi-local unitary, whose spatial
tails cannot generally be captured exactly by
the strictly local gate structure of conventional
MERA. To represent these tails, we introduce
auxiliary bonds connecting local gates, allowing
information to propagate over arbitrary distances
within each RG layer.

One possible realization is the fermionic Gaussian
unitary network introduced in Ref.~\cite{paperC},
which preserves global unitarity exactly by
construction. In that work, we developed a
scale-invariant layered architecture suited to
the repeated application of the downsampled
zipper unitary $Z$ considered here. The same local
tensor $u$ is used at every spatial position
and across all layers, making $u$ the sole
variational tensor. Further details of the
unitary network are provided in
Appendix~\ref{app:unitary_network}.

We variationally optimize the unitary network
$U_{\mathrm{Net}}[u]$ to approximate the
single-layer zipper unitary $Z$.
As shown in Fig.~\ref{fig:un_reconstruction},
the optimized network achieves high fidelity
with the target zipper unitary and exhibits
exponentially decaying real-space matrix elements.
However, when the same optimized network is
reused across RG scales, the reconstructed state
develops long-distance correlations consistent
with algebraic decay, rather than the exponential
decay of the target Chern-insulator state.
Increasing the bond dimension improves the
reconstruction fidelity and extends the range
over which the correlations agree with those
of the target state. Nevertheless, the apparent
algebraic tail persists, with its onset shifted
to larger separations. Thus, an accurate
single-layer approximation does not necessarily
guarantee faithful long-distance correlations
under scale-invariant reconstruction.

\section{\label{sec: discussion} Discussion: What makes a good renormalization group?}

The arbitrariness of the ZER fixed point originates from the large 
flexibility of the zipper unitary $Z$. The only requirements 
imposed on the zipper unitary $Z$ are:
\begin{enumerate}
    \item Fully disentangle $\frac{b^d-1}{b^d}$ of the modes from the system.
    \item Have an exponential decay tail.
\end{enumerate}
These two conditions do not fully constrain the form of $Z$. 
In particular, applying the transformation $S$ introduced in 
Subsection~\ref{subsec:A path to all other states} can generate a 
different disentangler $Z'$ that also satisfies these two requirements. 
Therefore, additional conditions may be needed to further constrain the 
RG procedure.

One possible direction is to require the zipper unitary to maximize 
the amount of disentanglement. After fully disentangling $\frac{b^d-1}{b^d}$ of the 
modes, there remains the possibility of further reducing the 
entanglement among the remaining modes. In fact, entanglement 
renormalization employs an optimization procedure to determine the 
optimal disentangler.

However, unlike entanglement renormalization, where the disentangler $u$
is given by local gates, the zipper unitary 
$Z$ is a more general nonlocal object with an exponential decay 
tail. Therefore, finding an appropriate parameterization of such 
exponentially decaying unitaries remains an open question.

Another possible direction is to require the zipper unitary $Z$ 
to be as local as possible. The motivation is that, since RG performs 
coarse graining, it is desirable to avoid unnecessary modifications of 
long-range correlations. Although the exponential decay structure of the 
zipper unitary $Z$ ensures locality in an asymptotic sense, 
its long-range tail can still introduce nonlocal contributions. 
Therefore, additional constraints can be imposed to suppress these 
effects. Along this line, one possible measure of locality is
\begin{equation}
    \sum_{R} ||U_{R}||^2 |R|^2 ,
\end{equation}
where $U_{R}$ is the Fourier transformation of $U(k)$. This quantity 
measures the quadratic spatial spread of the unitary transformation in 
real space and is analogous to the spread functional used in the 
construction of maximally localized Wannier functions \cite{Marzari1997Maximally}. 
Its square root defines a characteristic length scale of the unitary 
transformation and can be used as a measure of the locality of the 
zipper unitary $Z$.

These two directions may represent competing requirements and may not 
be simultaneously optimized. However, exploring the balance between 
maximal disentanglement and locality provides an interesting direction 
for future studies.

\section{\label{sec: conclusion} Conclusion}

We apply zipper entanglement renormalization (ZER)
to gapped free-fermion states, introducing a modified
construction of the disentangling zipper unitary that enables exact
state factorization. Unlike conventional entanglement
renormalization, ZER allows disentangling unitaries
with exponentially decaying spatial tails.
These transformations can accurately disentangle exponentially
localized occupied and empty orbitals. For Chern insulators, after a subset of modes is disentangled, the
nontrivial topology is retained in the remaining
courier degrees of freedom.

Applying ZER to Chern insulators, we find numerically
that the RG flow converges to an operational fixed
point under repeated iterations. Remarkably, the
resulting fixed-point state retains a finite,
nonzero correlation length, in contrast to the
vanishing or divergent correlation lengths commonly
associated with conventional RG fixed points.
We show that this behavior can be understood from
the quasi-local nature of the zipper unitary.
Furthermore, the ZER fixed point is not unique:
once a representative fixed-point pair is obtained, all other states within the same gapped phase can,
in principle, be realized as fixed points through
suitably chosen exponentially quasi-local
zipper unitaries.

The ZER fixed-point structure naturally leads to
a scale-invariant reconstruction of gapped
free-fermion states. Starting from a single-site
state $P_T=P(\tilde{\bm{k}}=\bm{0})$, the original
state $P_0$ is reconstructed by successively
introducing product-state sectors $P_{\mathrm{prod}}$
and applying inverse zipper unitaries. Scale invariance
follows from using the same momentum-dependent
zipper unitary at every RG scale, restricted to the
corresponding Brillouin-zone grid. This construction
is analogous to scale-invariant MERA but employs
exponentially quasi-local disentangling unitaries,
enabling state reconstruction with errors approaching
machine precision.

We find that the long-distance correlations of
reconstructed Chern-insulator states are highly
sensitive to perturbations. In particular, small
perturbations deliberately introduced into the
reconstruction can change the exponential decay
of correlations in the target state into algebraic
decay, even when the reconstructed state retains
high fidelity. The tensor-network representations also fail to reproduce the correct long-distance correlation behavior,
even when employing the layered architecture of ZER.

\textit{Note added:
While finalizing the present manuscript, Ref.~\cite{cirac2026efficientrepresentationspepscritical} appeared.  This recent preprint discussed how tensor network approximations to critical and gapped states could be obtained with bond dimensions that are polynomial in system size. Such discussions are related to our tensor-network approximation to the zipper unitaries, in that a polynomial scaling of overall bond dimension also emerges in our hierarchical architecture when the unitary network at each RG time has a fixed bond dimension. While we find that our tensor-network construction does not faithfully capture the exponential decay in the correlation functions, the behavior of correlation functions at a fixed state-approximation error was not explicitly discussed in Ref.~\cite{cirac2026efficientrepresentationspepscritical}.
}

\begin{acknowledgments}
This work is supported by the Hong Kong Research Grants Council (grants AoE/P-604/25-R, C7037-22G, 16308822, and 16310625), the Croucher Foundation (CIA23SC01), and the Future Science Awards Foundation through a Asian Young Scientist Fellowship. \\

ChatGPT (OpenAI; GPT-6) was used to assist with the generation of schematic diagrams, the development of numerical codes, and literature review. All AI-assisted visual materials and numerical implementations were critically reviewed, revised where necessary, and independently verified by the authors. The authors take full responsibility for the overall content of the manuscript.

\end{acknowledgments}

\appendix

\section{Quasi-local unitary with exponential decay}
\label{app:exponentially_decaying}

A unitary $U$ acting on a lattice many-body Hilbert space is
said to be quasi-local, or approximately locality-preserving,
with exponentially decaying tails
\cite{lieb1972finite,Ranard_2022} if there exist constants
$C,\mu>0$, independent of system size, such that
\begin{equation}
    \left\|[U^\dagger O_X U,O_Y]\right\|
    \leq
    C |X| |Y| \|O_X\| \|O_Y\|
    e^{-\mu d(X,Y)}
\end{equation}
for any local operators $O_X$ and $O_Y$ supported on
regions $X$ and $Y$, respectively.
Here, $d(X,Y)$ denotes the minimum distance between the
two regions, and $|X|$ and $|Y|$ denote their numbers of sites.
For fermionic systems, we restrict $O_X$ and $O_Y$ to
even-parity local operators.

For particle-number-conserving Gaussian unitaries,
quasi-locality can be characterized at the single-particle
level. Specifically, a single-particle unitary $u$ has
exponentially decaying matrix elements if there exist
constants $C,\mu>0$, independent of system size, such that
\begin{equation}
    \left|\langle x,\alpha|u|y,\beta\rangle\right|
    \leq C e^{-\mu d(x,y)}
\end{equation}
for all lattice sites $x,y$ and internal orbital indices
$\alpha,\beta$.

\section{L\"owdin orthonormalization}
\label{app:lowdin}

L\"owdin orthonormalization \cite{lowdin1950non} provides a systematic procedure for transforming a set of linearly independent, nonorthogonal basis states into an orthonormal set.

Consider a set of $N$ linearly independent basis states
$\{|\phi_i\rangle\}_{i=1}^{N}$, with the overlap matrix defined as
\begin{equation}
    S_{ij} = \langle \phi_i | \phi_j \rangle.
\end{equation}
Since the basis states are linearly independent, the overlap matrix
$S$ is Hermitian and positive definite. It can therefore be
diagonalized as
\begin{equation}
    S
    = U\Lambda U^{\dagger},
\end{equation}
where $U$ is unitary and
$\Lambda=\mathrm{diag}(\lambda_1,\ldots,\lambda_N)$
contains the positive eigenvalues of $S$. The inverse square root of the overlap matrix $S$ is defined by
\begin{equation}
    S^{-1/2}
    =
    U\Lambda^{-1/2}
    U^{\dagger},
\end{equation}
where
\begin{equation}
    \Lambda^{-1/2}
    =
    \mathrm{diag}
    \left(
        \lambda_1^{-1/2},\ldots,\lambda_N^{-1/2}
    \right).
\end{equation}
The L\"owdin-orthonormalized basis functions are then constructed as
\begin{equation}
    |\widetilde{\phi}_i\rangle
    =
    \sum_{j=1}^{N}
    |\phi_j\rangle
    \left(S^{-1/2}\right)_{ji}.
    \label{eq:lowdin_basis}
\end{equation}
Their orthonormality follows directly from
\begin{equation}
      \langle \widetilde{\phi}_i
    | \widetilde{\phi}_j \rangle
    =\left(
        S^{-1/2}
        S
        S^{-1/2}
    \right)_{ij} = \delta_{ij}.  
\end{equation}

If the overlap matrix $S$ has very small eigenvalues, the states
$\{|\phi_i\rangle\}_{i=1}^{N}$ are nearly linearly dependent,
making the computation of $S^{-1/2}$ numerically unstable.
To address this issue, we discard eigenvectors of $S$ whose
eigenvalues fall below a prescribed threshold, thereby removing
nearly linearly dependent directions and obtaining an orthonormal
basis of reduced dimension.

\section{Wannier interpolation}
\label{app:wannier_interpolation}

Wannier interpolation \cite{Marzari2012Maximally,Wang2006Abinitio}
provides an efficient framework for evaluating momentum-dependent
observables on dense momentum-space grids from their values on
coarser grids.

In the following discussion, we use the square lattice as an example, although the similar procedure can be generalized to other lattice geometries. Consider a momentum-dependent observable $O(\bm{\tilde{k}})$ defined on
the $L\times L$ dimensionless momentum grid $\Tilde{\mathrm{BZ}}_{\tau+1}$,
\begin{equation}
    \Tilde{\bm{k}}_{mn}
    = \frac{2\pi}{L}(m,n) \in \Tilde{\mathrm{BZ}}_{\tau+1},
    \qquad m,n=0,\ldots,L-1,
\end{equation}
where $L$ is a dimensionless integer. To interpolate $O(\tilde{\bm{k}})$ onto a finer $2L\times 2L$ grid $\Tilde{\mathrm{BZ}}_{\tau}$,
we first perform a discrete Fourier transformation to obtain
its real-space representation,
\begin{equation}
    O(\bm{\tilde{R}})
    =
    \frac{1}{L^2}
    \sum_{\bm{k}}
    e^{-i\bm{\tilde{k}}\cdot\bm{\tilde{R}}}
    O(\bm{\tilde{k}}),
    \label{eq:observable_fourier}
\end{equation}
where $\bm{\tilde{R}}=(\tilde{R}_x,\tilde{R}_y)$ denotes a lattice vector in the
centered real-space supercell, with
$\tilde{R}_x,\tilde{R}_y=-L/2,\ldots,L/2-1$ for even $L$.

The Fourier coefficients are then embedded into a
$2L\times 2L$ real-space array by placing them at the
center and padding the remaining entries with zeros.
The observable on the finer momentum grid is reconstructed
through the inverse Fourier transformation,
\begin{equation}
    \widetilde{O}(\bm{\tilde{q}})
    =
    \sum_{\bm{R}}
    e^{i\bm{\tilde{q}}\cdot\bm{\tilde{R}}}
    O(\bm{\tilde{R}}),
    \qquad
    \bm{\tilde{q}}=\frac{2\pi}{2L}(m,n) \in \Tilde{\mathrm{BZ}}_{\tau},
    \label{eq:observable_interpolation}
\end{equation}
where $m,n=0,\ldots,2L-1$ and the sum is taken over
the original real-space supercell.
The accuracy of this interpolation relies on the
localization of $O(\bm{\tilde{R}})$, which suppresses finite-size
effects associated with the truncation of its real-space
representation.

In our RG procedure, we apply this interpolation to the
momentum-space projector $P(\bm{\tilde{k}})$, obtaining
$\widetilde{P}(\bm{\tilde{q}})$ on the finer grid.
Although the interpolated matrix remains Hermitian,
it generally does not satisfy the projector condition
$\widetilde{P}^2=\widetilde{P}$ exactly.
We therefore restore the projector structure by
diagonalizing $\widetilde{P}(\bm{\tilde{q}})$ and retaining
the eigenvectors corresponding to its $N_{\mathrm{occ}}$
largest eigenvalues:
\begin{equation}
    P_{\mathrm{interp}}(\bm{\tilde{q}})
    =
    \sum_{n=1}^{N_{\mathrm{occ}}}
    |u_n(\bm{\tilde{q}})\rangle
    \langle u_n(\bm{\tilde{q}})|,
    \label{eq:projector_interpolation}
\end{equation}
where $N_{\mathrm{occ}}$ is the rank of the original
projector. This construction ensures that the upsampled
projector is Hermitian and idempotent, with the same rank
as the original projector.

\section{Kato parallel transport}
\label{app: Connecting}

In this appendix, we show that a path connecting two Bloch projectors
$P_0$ and $P_1$ in the same gapped phase induces an exponentially
localized unitary transformation between them. Let $P_s(\bm{k})$,
$s\in[0,1]$, be a family of rank-$r$ orthogonal projectors on the
Brillouin torus $\mathbb{T}^d$, smooth in $s$ and satisfying
\begin{equation}
    P_s(\bm{k})^2=P_s(\bm{k}),\qquad
    P_s(\bm{k})^\dagger=P_s(\bm{k}),
\end{equation}
with
\begin{equation}
    P_{s=0}(\bm{k})=P_0(\bm{k}),\qquad
    P_{s=1}(\bm{k})=P_1(\bm{k}).
\end{equation}

For a uniformly gapped family of quasi-local Bloch Hamiltonians whose
real-space matrix elements decay exponentially, the corresponding
spectral projectors are analytic in $\bm{k}$ in a complex neighborhood
of the Brillouin torus
\cite{des1964energy, Brouder2007Exponential}. We therefore assume that
$P_s(\bm{k})$ admits an analytic continuation to the strip
\begin{equation}
    \mathcal{S}_{\kappa}
    =
    \left\{
    \bm{k}\in\mathbb{C}^d:
    |\operatorname{Im}k_j|<\kappa,\quad j=1,\ldots,d
    \right\},
\end{equation}
with some $\kappa>0$ that can be chosen uniformly for
$s\in[0,1]$.

We connect the two projectors by Kato parallel transport
\cite{kato1950adiabatic}:
\begin{equation}
    \partial_s U_s(k)
    =
    A_s(k)U_s(k),
    \qquad
    U_0(k)=\mathbb I,
\end{equation}
with Kato generator
\begin{equation}
    A_s(k)
    =
    [\partial_s P_s(k),P_s(k)].
\end{equation}
On the real Brillouin torus, $A_s(k)$ is anti-Hermitian,
\begin{equation}
    A_s(k)^\dagger=-A_s(k),
\end{equation}
so the corresponding evolution $U_s(k)$ is unitary.

We now show that this evolution transports $P_0(k)$ into $P_s(k)$.
Differentiating the projector identity $P_s^2=P_s$ with respect to $s$
gives
\begin{equation}
    (\partial_sP_s)P_s+P_s(\partial_sP_s)
    =
    \partial_sP_s.
\end{equation}
Multiplying this identity by $P_s$ from either side yields
\begin{equation}
    P_s(\partial_sP_s)P_s=0.
\end{equation}
Consequently,
\begin{align}
    [A_s,P_s]
    &=
    [[\partial_sP_s,P_s],P_s]
    \nonumber\\
    &=
    (\partial_sP_s)P_s
    +P_s(\partial_sP_s)
    -2P_s(\partial_sP_s)P_s
    \nonumber\\
    &=
    \partial_sP_s.
\end{align}
Thus $P_s$ itself satisfies the evolution equation
\begin{equation}
    \partial_sP_s=[A_s,P_s].
\end{equation}
Hence $P_s$ and transported initial projector  $\widetilde{P}_s(k)
    =
    U_s(k)P_0(k)U_s(k)^\dagger$ satisfy the same first-order differential equation, with the same initial condition,
\begin{equation}
    \widetilde{P}_0=P_0.
\end{equation}
By uniqueness of the solution, they coincide:
\begin{equation}
    P_s(k)
    =
    U_s(k)P_0(k)U_s(k)^\dagger.
\end{equation}
Evaluating at $s=1$ gives
\begin{equation}
    P_1(k)
    =
    U(k)P_0(k)U(k)^\dagger,
    \qquad
    U(k)\equiv U_1(k).
\end{equation}

It remains to establish exponential locality. Since $P_s(k)$ is analytic in $\mathcal S_\kappa$, so is $
A_s(k)
=
[\partial_sP_s(k),P_s(k)]$. The solution can be represented by the convergent Dyson series
\begin{equation}
U_s(k)
=
\mathbb I+
\sum_{n=1}^{\infty}
\int_{0<s_1<\cdots<s_n<s}
A_{s_n}(k)\cdots A_{s_1}(k)
\,ds_1\cdots ds_n .
\end{equation}
Uniform boundedness of $A_s(k)$ on every closed substrip
$$
|\operatorname{Im}k_j|\leq\kappa'<\kappa
$$
implies uniform convergence of this series there. Hence $U_s(k)$, and in particular $U(k)=U_1(k)$, is analytic and bounded on every such substrip.

Define the real-space Fourier coefficients of $U(k)$ by
\begin{equation}
    U(R)
    =
    \frac{1}{|\mathrm{BZ}|}
    \int_{\mathrm{BZ}}
    d^dk\,
    e^{ik\cdot R}U(k).
\end{equation}
The exponential decay of $U(R)$ follows from the standard relation
between complex analyticity and the decay of Fourier coefficients \cite{des1964energy,Brouder2007Exponential}:
a periodic function that is analytic and bounded in the strip
\begin{equation}
    |\operatorname{Im} k_j|<\kappa
\end{equation}
has Fourier coefficients that decay exponentially at any rate
$\kappa'<\kappa$.

\section{Direct rotation between a pair of subspaces}
\label{app:geodesic_projector}

\begin{figure}
    \centering
    \includegraphics[width=0.8\linewidth]{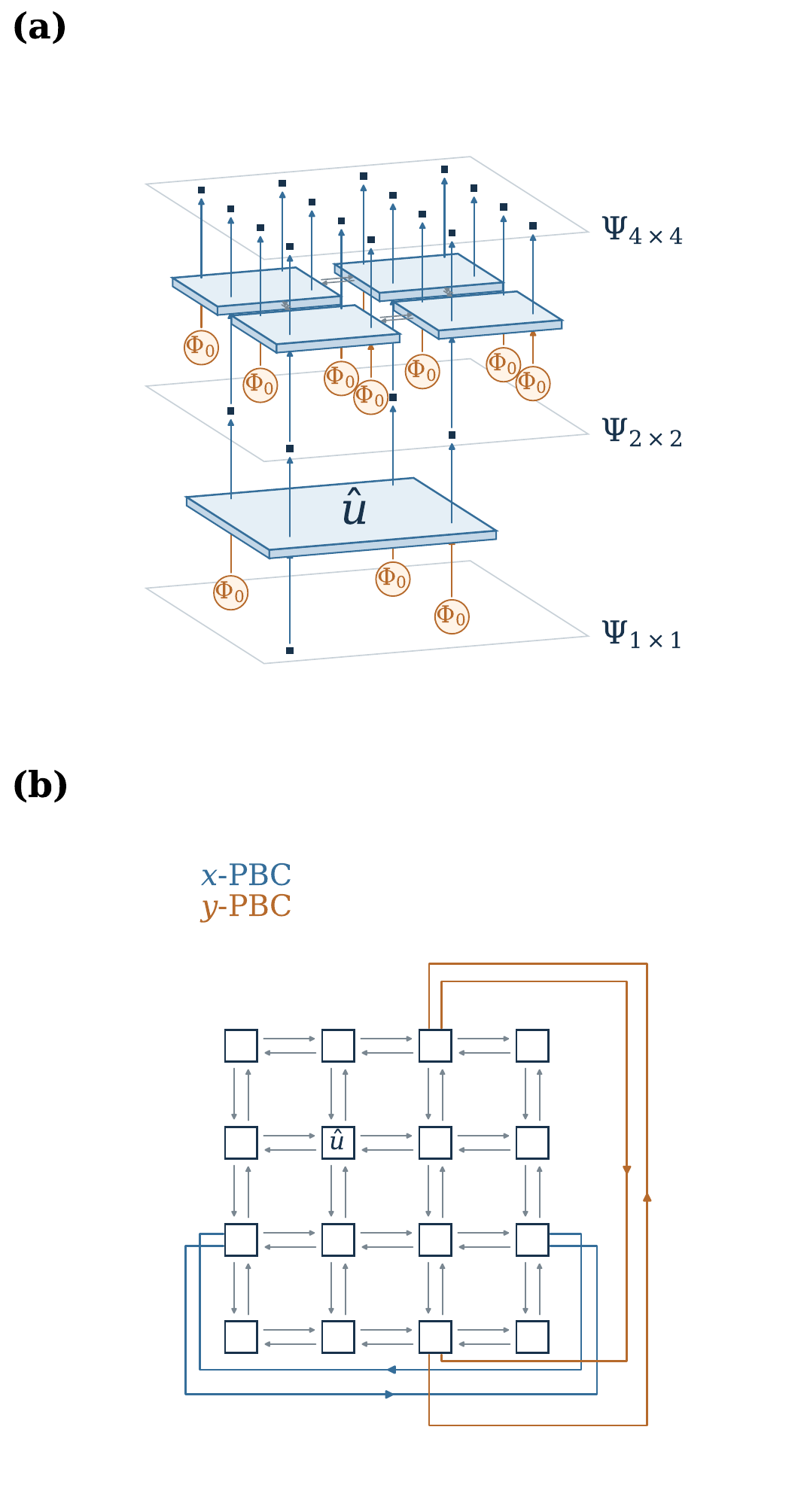}
    \caption{
    Fermionic Gaussian unitary network for hierarchical state
    reconstruction.
    (a) Successive network layers reconstruct $\Psi_{2\times2}$ and
    $\Psi_{4\times4}$ from the single-site state $\Psi_{1\times1}$.
    At each layer, product-state inputs $\Phi_{0}$ are added
    and then transformed by the unitary network
    $U_{\mathrm{Net}}[u]$.
    (b) Spatial connectivity within a single layer of the unitary network,
    with periodic boundary conditions imposed in both spatial directions. The schematic diagram was generated with assistance from ChatGPT (OpenAI; GPT-6) and subsequently reviewed and verified by the authors.
}
    \label{fig:unitary_network}
\end{figure}

In \cite{Bravyi_2011}, the direct rotation between two subspaces described by projectors $P_0$ and $P_1$ is introduced. This construction can be seens as the particle-number-conserving counterpart of the geodesic construction for fermionic Gaussian states in \cite{Hackl_2018}. For particle-number-conserving case, the state manifold is the complex Grassmannian, and the direct rotation gives the minimal geodesic away from the cut locus.

Define the reflection operators
\begin{equation}
R_i
=
1-2P_i,
\qquad
R_i^\dagger=R_i,
\qquad
R_i^2=1,
\qquad i=0,1.
\end{equation}
Their product
\begin{equation}
    \Delta = R_1R_0
\end{equation}
plays the role of the relative covariance matrix of
\cite{Hackl_2018}. Since $R_i^2=1$, $\Delta$ is unitary, $ \Delta^\dagger=\Delta^{-1} $.
If no nonzero vector in $\operatorname{Ran}P_0$ is orthogonal to
$\operatorname{Ran}P_1$, and vice versa, then
$-1\notin\operatorname{spec}\Delta$. In this case, the direct rotation is
well defined by
\begin{equation}
    U=\Delta^{1/2}=\exp{K},
\end{equation}
where
\[
 K=\frac{1}{2}\log\Delta,
\]
and the logarithm is taken with eigenphases in $(-\pi,\pi)$. The associated geodesic is
\begin{equation}
    P_s=U_sP_0U_s^\dagger, \qquad
    U_s=e^{sK},
    \qquad
    0\leq s\leq 1,
\end{equation}
To verify the endpoint, note that
\begin{equation}
    R_0\Delta R_0=\Delta^{-1},
\end{equation}
which implies
\begin{equation}
    R_0KR_0=-K.
\end{equation}
Therefore,
\begin{equation}
UR_0U^\dagger
=
e^KR_0e^{-K}
=
e^{2K}R_0
=
\Delta R_0
=
R_1.
\end{equation}
Using $P_i=(1-R_i)/2$, we obtain
\begin{equation}
    P_1=UP_0U^\dagger.
\end{equation}

To apply this construction to translation-invariant,
particle-number-conserving fermionic Gaussian states,
\[
 P_i=\bigoplus_{\bm{k}}P_i(\bm{k}),
    \qquad i=0,1.
\]
the direct rotation can be constructed independently for each crystal momentum sector $\bm{k}$. Define the reflection operators
\begin{equation}
R_i(\bm{k})
=
1-2P_i(\bm{k}),
\end{equation}
and the relative covariance matrices
\begin{equation}
\Delta(\bm{k})
=
R_1(\bm{k})R_0(\bm{k}).
\label{eq:relative_projector}
\end{equation}
The direct rotation is given by
\begin{equation}
P_1(\bm{k})
=
U(\bm{k})
P_0(\bm{k})
U^\dagger(\bm{k}),  \quad U(\bm{k}) = \Delta^{1/2}(\bm{k})
\end{equation}
If $P_0(\bm{k})$ and $P_1(\bm{k})$ are smooth (or analytic) in $\bm{k}$ and
\begin{equation}
-1\notin\operatorname{spec}\Delta(\bm{k})
\end{equation}
throughout the Brillouin zone, then  $U(\bm{k})$ inherit the same smoothness (or analyticity). If the spectrum
crosses $-1$, the calculation of $U(\bm{k})$ requires an additional branch choice.

\section{Unitary-network ansatz for ZER reconstruction}
\label{app:unitary_network}

To represent the zipper unitary as a tensor-network unitary
ansatz, we use the fermionic Gaussian unitary network introduced in
Ref.~\cite{paperC}. This construction provides a parametrization of a
broad class of global Gaussian unitaries, including quasi-local
unitaries with exponentially decaying tails. Global unitarity is
guaranteed by construction, so no additional constraints need to be
imposed on the variational parameters. Tensor contractions can be
performed at the single-particle level.

Ref.~\cite{paperC} introduced a scale-invariant layered architecture
that is naturally suited to the scale-invariant state reconstruction
considered here. For the Chern-insulator state reconstruction, the network is illustrated in Fig.~\ref{fig:unitary_network}. At each layer, product-state inputs $\Phi_{0}$ are added and subsequently transformed by one layer of the unitary network $U_{\mathrm{Net}}[u]$. We variationally optimize a single network layer $U_{\mathrm{Net}}[u]$ to approximate the zipper unitary $Z$. 

At each RG scale, the zipper unitary $Z$ is translation invariant with respect to the coarse-grained lattice and, under downsampling, is described by the same momentum-dependent
function across scales. Correspondingly, in the unitary-network ansatz,
we use the same local tensor $u$ at every spatial position and in every
layer, so that $u$ constitutes the only set of variational parameters. By stacking these layers and successively adding the
product-state sectors, the network reconstructs the target
Chern-insulator state from the single-site state.

\bibliography{ref}

\begin{thebibliography}{50}%
\makeatletter
\providecommand \@ifxundefined [1]{%
 \@ifx{#1\undefined}
}%
\providecommand \@ifnum [1]{%
 \ifnum #1\expandafter \@firstoftwo
 \else \expandafter \@secondoftwo
 \fi
}%
\providecommand \@ifx [1]{%
 \ifx #1\expandafter \@firstoftwo
 \else \expandafter \@secondoftwo
 \fi
}%
\providecommand \natexlab [1]{#1}%
\providecommand \enquote  [1]{``#1''}%
\providecommand \bibnamefont  [1]{#1}%
\providecommand \bibfnamefont [1]{#1}%
\providecommand \citenamefont [1]{#1}%
\providecommand \href@noop [0]{\@secondoftwo}%
\providecommand \href [0]{\begingroup \@sanitize@url \@href}%
\providecommand \@href[1]{\@@startlink{#1}\@@href}%
\providecommand \@@href[1]{\endgroup#1\@@endlink}%
\providecommand \@sanitize@url [0]{\catcode `\\12\catcode `\$12\catcode
  `\&12\catcode `\#12\catcode `\^12\catcode `\_12\catcode `\%12\relax}%
\providecommand \@@startlink[1]{}%
\providecommand \@@endlink[0]{}%
\providecommand \url  [0]{\begingroup\@sanitize@url \@url }%
\providecommand \@url [1]{\endgroup\@href {#1}{\urlprefix }}%
\providecommand \urlprefix  [0]{URL }%
\providecommand \Eprint [0]{\href }%
\providecommand \doibase [0]{https://doi.org/}%
\providecommand \selectlanguage [0]{\@gobble}%
\providecommand \bibinfo  [0]{\@secondoftwo}%
\providecommand \bibfield  [0]{\@secondoftwo}%
\providecommand \translation [1]{[#1]}%
\providecommand \BibitemOpen [0]{}%
\providecommand \bibitemStop [0]{}%
\providecommand \bibitemNoStop [0]{.\EOS\space}%
\providecommand \EOS [0]{\spacefactor3000\relax}%
\providecommand \BibitemShut  [1]{\csname bibitem#1\endcsname}%
\let\auto@bib@innerbib\@empty
\bibitem [{\citenamefont {Fisher}(1998)}]{Fisher1998Renormalization}%
  \BibitemOpen
  \bibfield  {author} {\bibinfo {author} {\bibfnamefont {M.~E.}\ \bibnamefont
  {Fisher}},\ }\bibfield  {title} {\bibinfo {title} {Renormalization group
  theory: Its basis and formulation in statistical physics},\ }\href
  {https://doi.org/10.1103/RevModPhys.70.653} {\bibfield  {journal} {\bibinfo
  {journal} {Rev. Mod. Phys.}\ }\textbf {\bibinfo {volume} {70}},\ \bibinfo
  {pages} {653} (\bibinfo {year} {1998})}\BibitemShut {NoStop}%
\bibitem [{\citenamefont {Wilson}(1975)}]{Wilson1975renormalization}%
  \BibitemOpen
  \bibfield  {author} {\bibinfo {author} {\bibfnamefont {K.~G.}\ \bibnamefont
  {Wilson}},\ }\bibfield  {title} {\bibinfo {title} {The renormalization group:
  Critical phenomena and the kondo problem},\ }\href
  {https://doi.org/10.1103/RevModPhys.47.773} {\bibfield  {journal} {\bibinfo
  {journal} {Rev. Mod. Phys.}\ }\textbf {\bibinfo {volume} {47}},\ \bibinfo
  {pages} {773} (\bibinfo {year} {1975})}\BibitemShut {NoStop}%
\bibitem [{\citenamefont {White}(1992)}]{White1992Density}%
  \BibitemOpen
  \bibfield  {author} {\bibinfo {author} {\bibfnamefont {S.~R.}\ \bibnamefont
  {White}},\ }\bibfield  {title} {\bibinfo {title} {Density matrix formulation
  for quantum renormalization groups},\ }\href
  {https://doi.org/10.1103/PhysRevLett.69.2863} {\bibfield  {journal} {\bibinfo
   {journal} {Phys. Rev. Lett.}\ }\textbf {\bibinfo {volume} {69}},\ \bibinfo
  {pages} {2863} (\bibinfo {year} {1992})}\BibitemShut {NoStop}%
\bibitem [{\citenamefont {Verstraete}\ \emph {et~al.}(2004)\citenamefont
  {Verstraete}, \citenamefont {Porras},\ and\ \citenamefont
  {Cirac}}]{Verstraete2004Density}%
  \BibitemOpen
  \bibfield  {author} {\bibinfo {author} {\bibfnamefont {F.}~\bibnamefont
  {Verstraete}}, \bibinfo {author} {\bibfnamefont {D.}~\bibnamefont {Porras}},\
  and\ \bibinfo {author} {\bibfnamefont {J.~I.}\ \bibnamefont {Cirac}},\
  }\bibfield  {title} {\bibinfo {title} {Density matrix renormalization group
  and periodic boundary conditions: A quantum information perspective},\ }\href
  {https://doi.org/10.1103/PhysRevLett.93.227205} {\bibfield  {journal}
  {\bibinfo  {journal} {Phys. Rev. Lett.}\ }\textbf {\bibinfo {volume} {93}},\
  \bibinfo {pages} {227205} (\bibinfo {year} {2004})}\BibitemShut {NoStop}%
\bibitem [{\citenamefont {Verstraete}\ \emph {et~al.}(2008)\citenamefont
  {Verstraete}, \citenamefont {Murg},\ and\ \citenamefont
  {Cirac}}]{Verstraete_2008}%
  \BibitemOpen
  \bibfield  {author} {\bibinfo {author} {\bibfnamefont {F.}~\bibnamefont
  {Verstraete}}, \bibinfo {author} {\bibfnamefont {V.}~\bibnamefont {Murg}},\
  and\ \bibinfo {author} {\bibfnamefont {J.}~\bibnamefont {Cirac}},\ }\bibfield
   {title} {\bibinfo {title} {Matrix product states, projected entangled pair
  states, and variational renormalization group methods for quantum spin
  systems},\ }\href {https://doi.org/10.1080/14789940801912366} {\bibfield
  {journal} {\bibinfo  {journal} {Advances in Physics}\ }\textbf {\bibinfo
  {volume} {57}},\ \bibinfo {pages} {143–224} (\bibinfo {year}
  {2008})}\BibitemShut {NoStop}%
\bibitem [{\citenamefont {Cirac}\ \emph {et~al.}(2021)\citenamefont {Cirac},
  \citenamefont {P{\'e}rez-Garc{\'{\i}}a}, \citenamefont {Schuch},\ and\
  \citenamefont {Verstraete}}]{Cirac2021Matrix}%
  \BibitemOpen
  \bibfield  {author} {\bibinfo {author} {\bibfnamefont {J.~I.}\ \bibnamefont
  {Cirac}}, \bibinfo {author} {\bibfnamefont {D.}~\bibnamefont
  {P{\'e}rez-Garc{\'{\i}}a}}, \bibinfo {author} {\bibfnamefont
  {N.}~\bibnamefont {Schuch}},\ and\ \bibinfo {author} {\bibfnamefont
  {F.}~\bibnamefont {Verstraete}},\ }\bibfield  {title} {\bibinfo {title}
  {Matrix product states and projected entangled pair states: Concepts,
  symmetries, theorems},\ }\href {https://doi.org/10.1103/RevModPhys.93.045003}
  {\bibfield  {journal} {\bibinfo  {journal} {Rev. Mod. Phys.}\ }\textbf
  {\bibinfo {volume} {93}},\ \bibinfo {pages} {045003} (\bibinfo {year}
  {2021})}\BibitemShut {NoStop}%
\bibitem [{\citenamefont {Schollwöck}(2011)}]{Schollw_ck_2011}%
  \BibitemOpen
  \bibfield  {author} {\bibinfo {author} {\bibfnamefont {U.}~\bibnamefont
  {Schollwöck}},\ }\bibfield  {title} {\bibinfo {title} {The density-matrix
  renormalization group in the age of matrix product states},\ }\href
  {https://doi.org/10.1016/j.aop.2010.09.012} {\bibfield  {journal} {\bibinfo
  {journal} {Annals of Physics}\ }\textbf {\bibinfo {volume} {326}},\ \bibinfo
  {pages} {96–192} (\bibinfo {year} {2011})}\BibitemShut {NoStop}%
\bibitem [{\citenamefont {Vidal}(2007)}]{Vidal2007Entanglement}%
  \BibitemOpen
  \bibfield  {author} {\bibinfo {author} {\bibfnamefont {G.}~\bibnamefont
  {Vidal}},\ }\bibfield  {title} {\bibinfo {title} {Entanglement
  renormalization},\ }\href {https://doi.org/10.1103/PhysRevLett.99.220405}
  {\bibfield  {journal} {\bibinfo  {journal} {Phys. Rev. Lett.}\ }\textbf
  {\bibinfo {volume} {99}},\ \bibinfo {pages} {220405} (\bibinfo {year}
  {2007})}\BibitemShut {NoStop}%
\bibitem [{\citenamefont {Vidal}(2008)}]{Vidal2008Class}%
  \BibitemOpen
  \bibfield  {author} {\bibinfo {author} {\bibfnamefont {G.}~\bibnamefont
  {Vidal}},\ }\bibfield  {title} {\bibinfo {title} {Class of quantum many-body
  states that can be efficiently simulated},\ }\href
  {https://doi.org/10.1103/PhysRevLett.101.110501} {\bibfield  {journal}
  {\bibinfo  {journal} {Phys. Rev. Lett.}\ }\textbf {\bibinfo {volume} {101}},\
  \bibinfo {pages} {110501} (\bibinfo {year} {2008})}\BibitemShut {NoStop}%
\bibitem [{\citenamefont
  {Vidal}(2010)}]{vidal2010entanglementrenormalizationintroduction}%
  \BibitemOpen
  \bibfield  {author} {\bibinfo {author} {\bibfnamefont {G.}~\bibnamefont
  {Vidal}},\ }\href {https://arxiv.org/abs/0912.1651} {\bibinfo {title}
  {Entanglement renormalization: an introduction}} (\bibinfo {year} {2010}),\
  \Eprint {https://arxiv.org/abs/0912.1651} {arXiv:0912.1651 [cond-mat.str-el]}
  \BibitemShut {NoStop}%
\bibitem [{\citenamefont {Pfeifer}\ \emph {et~al.}(2009)\citenamefont
  {Pfeifer}, \citenamefont {Evenbly},\ and\ \citenamefont
  {Vidal}}]{Pfeifer_2009}%
  \BibitemOpen
  \bibfield  {author} {\bibinfo {author} {\bibfnamefont {R.~N.~C.}\
  \bibnamefont {Pfeifer}}, \bibinfo {author} {\bibfnamefont {G.}~\bibnamefont
  {Evenbly}},\ and\ \bibinfo {author} {\bibfnamefont {G.}~\bibnamefont
  {Vidal}},\ }\bibfield  {title} {\bibinfo {title} {Entanglement
  renormalization, scale invariance, and quantum criticality},\ }\bibfield
  {journal} {\bibinfo  {journal} {Physical Review A}\ }\textbf {\bibinfo
  {volume} {79}},\ \href {https://doi.org/10.1103/physreva.79.040301}
  {10.1103/physreva.79.040301} (\bibinfo {year} {2009})\BibitemShut {NoStop}%
\bibitem [{\citenamefont {Evenbly}\ and\ \citenamefont
  {Vidal}(2010)}]{Evenbly2010Entanglement}%
  \BibitemOpen
  \bibfield  {author} {\bibinfo {author} {\bibfnamefont {G.}~\bibnamefont
  {Evenbly}}\ and\ \bibinfo {author} {\bibfnamefont {G.}~\bibnamefont
  {Vidal}},\ }\bibfield  {title} {\bibinfo {title} {Entanglement
  renormalization in noninteracting fermionic systems},\ }\href
  {https://doi.org/10.1103/PhysRevB.81.235102} {\bibfield  {journal} {\bibinfo
  {journal} {Phys. Rev. B}\ }\textbf {\bibinfo {volume} {81}},\ \bibinfo
  {pages} {235102} (\bibinfo {year} {2010})}\BibitemShut {NoStop}%
\bibitem [{\citenamefont {Evenbly}\ \emph {et~al.}(2010)\citenamefont
  {Evenbly}, \citenamefont {Pfeifer}, \citenamefont {Pic{\'o}}, \citenamefont
  {Iblisdir}, \citenamefont {Tagliacozzo}, \citenamefont {McCulloch},\ and\
  \citenamefont {Vidal}}]{Evenbly2010Boundary}%
  \BibitemOpen
  \bibfield  {author} {\bibinfo {author} {\bibfnamefont {G.}~\bibnamefont
  {Evenbly}}, \bibinfo {author} {\bibfnamefont {R.~N.~C.}\ \bibnamefont
  {Pfeifer}}, \bibinfo {author} {\bibfnamefont {V.}~\bibnamefont {Pic{\'o}}},
  \bibinfo {author} {\bibfnamefont {S.}~\bibnamefont {Iblisdir}}, \bibinfo
  {author} {\bibfnamefont {L.}~\bibnamefont {Tagliacozzo}}, \bibinfo {author}
  {\bibfnamefont {I.~P.}\ \bibnamefont {McCulloch}},\ and\ \bibinfo {author}
  {\bibfnamefont {G.}~\bibnamefont {Vidal}},\ }\bibfield  {title} {\bibinfo
  {title} {Boundary quantum critical phenomena with entanglement
  renormalization},\ }\href {https://doi.org/10.1103/PhysRevB.82.161107}
  {\bibfield  {journal} {\bibinfo  {journal} {Phys. Rev. B}\ }\textbf {\bibinfo
  {volume} {82}},\ \bibinfo {pages} {161107(R)} (\bibinfo {year}
  {2010})}\BibitemShut {NoStop}%
\bibitem [{\citenamefont {Corboz}\ \emph {et~al.}(2010)\citenamefont {Corboz},
  \citenamefont {Evenbly}, \citenamefont {Verstraete},\ and\ \citenamefont
  {Vidal}}]{Corboz2010Simulation}%
  \BibitemOpen
  \bibfield  {author} {\bibinfo {author} {\bibfnamefont {P.}~\bibnamefont
  {Corboz}}, \bibinfo {author} {\bibfnamefont {G.}~\bibnamefont {Evenbly}},
  \bibinfo {author} {\bibfnamefont {F.}~\bibnamefont {Verstraete}},\ and\
  \bibinfo {author} {\bibfnamefont {G.}~\bibnamefont {Vidal}},\ }\bibfield
  {title} {\bibinfo {title} {Simulation of interacting fermions with
  entanglement renormalization},\ }\href
  {https://doi.org/10.1103/PhysRevA.81.010303} {\bibfield  {journal} {\bibinfo
  {journal} {Phys. Rev. A}\ }\textbf {\bibinfo {volume} {81}},\ \bibinfo
  {pages} {010303(R)} (\bibinfo {year} {2010})}\BibitemShut {NoStop}%
\bibitem [{\citenamefont {Silvi}\ \emph {et~al.}(2010)\citenamefont {Silvi},
  \citenamefont {Giovannetti}, \citenamefont {Calabrese}, \citenamefont
  {Santoro},\ and\ \citenamefont {Fazio}}]{Silvi_2010}%
  \BibitemOpen
  \bibfield  {author} {\bibinfo {author} {\bibfnamefont {P.}~\bibnamefont
  {Silvi}}, \bibinfo {author} {\bibfnamefont {V.}~\bibnamefont {Giovannetti}},
  \bibinfo {author} {\bibfnamefont {P.}~\bibnamefont {Calabrese}}, \bibinfo
  {author} {\bibfnamefont {G.~E.}\ \bibnamefont {Santoro}},\ and\ \bibinfo
  {author} {\bibfnamefont {R.}~\bibnamefont {Fazio}},\ }\bibfield  {title}
  {\bibinfo {title} {Entanglement renormalization and boundary critical
  phenomena},\ }\href {https://doi.org/10.1088/1742-5468/2010/03/l03001}
  {\bibfield  {journal} {\bibinfo  {journal} {Journal of Statistical Mechanics:
  Theory and Experiment}\ }\textbf {\bibinfo {volume} {2010}},\ \bibinfo
  {pages} {L03001} (\bibinfo {year} {2010})}\BibitemShut {NoStop}%
\bibitem [{\citenamefont {Evenbly}\ and\ \citenamefont
  {Vidal}(2013)}]{evenbly2013quantumcriticalitymultiscaleentanglement}%
  \BibitemOpen
  \bibfield  {author} {\bibinfo {author} {\bibfnamefont {G.}~\bibnamefont
  {Evenbly}}\ and\ \bibinfo {author} {\bibfnamefont {G.}~\bibnamefont
  {Vidal}},\ }\href {https://arxiv.org/abs/1109.5334} {\bibinfo {title}
  {Quantum criticality with the multi-scale entanglement renormalization
  ansatz}} (\bibinfo {year} {2013}),\ \Eprint {https://arxiv.org/abs/1109.5334}
  {arXiv:1109.5334 [quant-ph]} \BibitemShut {NoStop}%
\bibitem [{\citenamefont {Evenbly}\ and\ \citenamefont
  {Vidal}(2014)}]{Evenbly2014Class}%
  \BibitemOpen
  \bibfield  {author} {\bibinfo {author} {\bibfnamefont {G.}~\bibnamefont
  {Evenbly}}\ and\ \bibinfo {author} {\bibfnamefont {G.}~\bibnamefont
  {Vidal}},\ }\bibfield  {title} {\bibinfo {title} {Class of highly entangled
  many-body states that can be efficiently simulated},\ }\href
  {https://doi.org/10.1103/PhysRevLett.112.240502} {\bibfield  {journal}
  {\bibinfo  {journal} {Phys. Rev. Lett.}\ }\textbf {\bibinfo {volume} {112}},\
  \bibinfo {pages} {240502} (\bibinfo {year} {2014})}\BibitemShut {NoStop}%
\bibitem [{\citenamefont {Evenbly}\ and\ \citenamefont
  {Vidal}(2015)}]{Evenbly2015Tensor}%
  \BibitemOpen
  \bibfield  {author} {\bibinfo {author} {\bibfnamefont {G.}~\bibnamefont
  {Evenbly}}\ and\ \bibinfo {author} {\bibfnamefont {G.}~\bibnamefont
  {Vidal}},\ }\bibfield  {title} {\bibinfo {title} {Tensor network
  renormalization yields the multiscale entanglement renormalization ansatz},\
  }\href {https://doi.org/10.1103/PhysRevLett.115.200401} {\bibfield  {journal}
  {\bibinfo  {journal} {Phys. Rev. Lett.}\ }\textbf {\bibinfo {volume} {115}},\
  \bibinfo {pages} {200401} (\bibinfo {year} {2015})}\BibitemShut {NoStop}%
\bibitem [{\citenamefont {Barthel}\ \emph {et~al.}(2010)\citenamefont
  {Barthel}, \citenamefont {Kliesch},\ and\ \citenamefont
  {Eisert}}]{Barthel_2010}%
  \BibitemOpen
  \bibfield  {author} {\bibinfo {author} {\bibfnamefont {T.}~\bibnamefont
  {Barthel}}, \bibinfo {author} {\bibfnamefont {M.}~\bibnamefont {Kliesch}},\
  and\ \bibinfo {author} {\bibfnamefont {J.}~\bibnamefont {Eisert}},\
  }\bibfield  {title} {\bibinfo {title} {Real-space renormalization yields
  finite correlations},\ }\bibfield  {journal} {\bibinfo  {journal} {Physical
  Review Letters}\ }\textbf {\bibinfo {volume} {105}},\ \href
  {https://doi.org/10.1103/physrevlett.105.010502}
  {10.1103/physrevlett.105.010502} (\bibinfo {year} {2010})\BibitemShut
  {NoStop}%
\bibitem [{\citenamefont {Wahl}\ \emph {et~al.}(2013)\citenamefont {Wahl},
  \citenamefont {Tu}, \citenamefont {Schuch},\ and\ \citenamefont
  {Cirac}}]{Wahl_2013}%
  \BibitemOpen
  \bibfield  {author} {\bibinfo {author} {\bibfnamefont {T.~B.}\ \bibnamefont
  {Wahl}}, \bibinfo {author} {\bibfnamefont {H.-H.}\ \bibnamefont {Tu}},
  \bibinfo {author} {\bibfnamefont {N.}~\bibnamefont {Schuch}},\ and\ \bibinfo
  {author} {\bibfnamefont {J.~I.}\ \bibnamefont {Cirac}},\ }\bibfield  {title}
  {\bibinfo {title} {Projected entangled-pair states can describe chiral
  topological states},\ }\bibfield  {journal} {\bibinfo  {journal} {Physical
  Review Letters}\ }\textbf {\bibinfo {volume} {111}},\ \href
  {https://doi.org/10.1103/physrevlett.111.236805}
  {10.1103/physrevlett.111.236805} (\bibinfo {year} {2013})\BibitemShut
  {NoStop}%
\bibitem [{\citenamefont {Niu}\ \emph {et~al.}(2024)\citenamefont {Niu},
  \citenamefont {Li}, \citenamefont {Chen},\ and\ \citenamefont
  {Poilblanc}}]{Niu_2024}%
  \BibitemOpen
  \bibfield  {author} {\bibinfo {author} {\bibfnamefont {S.}~\bibnamefont
  {Niu}}, \bibinfo {author} {\bibfnamefont {J.-W.}\ \bibnamefont {Li}},
  \bibinfo {author} {\bibfnamefont {J.-Y.}\ \bibnamefont {Chen}},\ and\
  \bibinfo {author} {\bibfnamefont {D.}~\bibnamefont {Poilblanc}},\ }\bibfield
  {title} {\bibinfo {title} {Chiral spin liquids with projected gaussian
  fermionic entangled pair states},\ }\bibfield  {journal} {\bibinfo  {journal}
  {Physical Review B}\ }\textbf {\bibinfo {volume} {109}},\ \href
  {https://doi.org/10.1103/physrevb.109.l081107} {10.1103/physrevb.109.l081107}
  (\bibinfo {year} {2024})\BibitemShut {NoStop}%
\bibitem [{\citenamefont {Wong}\ and\ \citenamefont
  {Potter}(2025)}]{wong2025entanglementrenormalizationcircuits2d}%
  \BibitemOpen
  \bibfield  {author} {\bibinfo {author} {\bibfnamefont {S.~L.}\ \bibnamefont
  {Wong}}\ and\ \bibinfo {author} {\bibfnamefont {A.~C.}\ \bibnamefont
  {Potter}},\ }\href {https://arxiv.org/abs/2506.04200} {\bibinfo {title}
  {Entanglement renormalization circuits for $2d$ gaussian fermion states}}
  (\bibinfo {year} {2025}),\ \Eprint {https://arxiv.org/abs/2506.04200}
  {arXiv:2506.04200 [quant-ph]} \BibitemShut {NoStop}%
\bibitem [{\citenamefont {Dubail}\ and\ \citenamefont
  {Read}(2015)}]{Dubail_2015}%
  \BibitemOpen
  \bibfield  {author} {\bibinfo {author} {\bibfnamefont {J.}~\bibnamefont
  {Dubail}}\ and\ \bibinfo {author} {\bibfnamefont {N.}~\bibnamefont {Read}},\
  }\bibfield  {title} {\bibinfo {title} {Tensor network trial states for chiral
  topological phases in two dimensions and a no-go theorem in any dimension},\
  }\bibfield  {journal} {\bibinfo  {journal} {Physical Review B}\ }\textbf
  {\bibinfo {volume} {92}},\ \href {https://doi.org/10.1103/physrevb.92.205307}
  {10.1103/physrevb.92.205307} (\bibinfo {year} {2015})\BibitemShut {NoStop}%
\bibitem [{\citenamefont {Li}\ and\ \citenamefont {Mong}(2019)}]{Li_2019}%
  \BibitemOpen
  \bibfield  {author} {\bibinfo {author} {\bibfnamefont {Z.}~\bibnamefont
  {Li}}\ and\ \bibinfo {author} {\bibfnamefont {R.~S.~K.}\ \bibnamefont
  {Mong}},\ }\bibfield  {title} {\bibinfo {title} {Entanglement renormalization
  for chiral topological phases},\ }\bibfield  {journal} {\bibinfo  {journal}
  {Physical Review B}\ }\textbf {\bibinfo {volume} {99}},\ \href
  {https://doi.org/10.1103/physrevb.99.241105} {10.1103/physrevb.99.241105}
  (\bibinfo {year} {2019})\BibitemShut {NoStop}%
\bibitem [{\citenamefont {Wong}\ \emph {et~al.}(2022)\citenamefont {Wong},
  \citenamefont {Pang},\ and\ \citenamefont
  {Po}}]{wong2022zipperentanglementrenormalizationfree}%
  \BibitemOpen
  \bibfield  {author} {\bibinfo {author} {\bibfnamefont {S.~L.}\ \bibnamefont
  {Wong}}, \bibinfo {author} {\bibfnamefont {K.~C.}\ \bibnamefont {Pang}},\
  and\ \bibinfo {author} {\bibfnamefont {H.~C.}\ \bibnamefont {Po}},\ }\href
  {https://arxiv.org/abs/2206.11761} {\bibinfo {title} {Zipper entanglement
  renormalization for free fermions}} (\bibinfo {year} {2022}),\ \Eprint
  {https://arxiv.org/abs/2206.11761} {arXiv:2206.11761 [quant-ph]} \BibitemShut
  {NoStop}%
\bibitem [{\citenamefont {Fishman}\ and\ \citenamefont
  {White}(2015)}]{Fishman2015Compression}%
  \BibitemOpen
  \bibfield  {author} {\bibinfo {author} {\bibfnamefont {M.~T.}\ \bibnamefont
  {Fishman}}\ and\ \bibinfo {author} {\bibfnamefont {S.~R.}\ \bibnamefont
  {White}},\ }\bibfield  {title} {\bibinfo {title} {Compression of correlation
  matrices and an efficient method for forming matrix product states of
  fermionic gaussian states},\ }\href
  {https://doi.org/10.1103/PhysRevB.92.075132} {\bibfield  {journal} {\bibinfo
  {journal} {Phys. Rev. B}\ }\textbf {\bibinfo {volume} {92}},\ \bibinfo
  {pages} {075132} (\bibinfo {year} {2015})}\BibitemShut {NoStop}%
\bibitem [{\citenamefont {Evenbly}\ and\ \citenamefont
  {White}(2016)}]{Evenbly2016Entanglement}%
  \BibitemOpen
  \bibfield  {author} {\bibinfo {author} {\bibfnamefont {G.}~\bibnamefont
  {Evenbly}}\ and\ \bibinfo {author} {\bibfnamefont {S.~R.}\ \bibnamefont
  {White}},\ }\bibfield  {title} {\bibinfo {title} {Entanglement
  renormalization and wavelets},\ }\href
  {https://doi.org/10.1103/PhysRevLett.116.140403} {\bibfield  {journal}
  {\bibinfo  {journal} {Phys. Rev. Lett.}\ }\textbf {\bibinfo {volume} {116}},\
  \bibinfo {pages} {140403} (\bibinfo {year} {2016})}\BibitemShut {NoStop}%
\bibitem [{\citenamefont {Evenbly}\ and\ \citenamefont
  {White}(2018)}]{Evenbly2018Representation}%
  \BibitemOpen
  \bibfield  {author} {\bibinfo {author} {\bibfnamefont {G.}~\bibnamefont
  {Evenbly}}\ and\ \bibinfo {author} {\bibfnamefont {S.~R.}\ \bibnamefont
  {White}},\ }\bibfield  {title} {\bibinfo {title} {Representation and design
  of wavelets using unitary circuits},\ }\href
  {https://doi.org/10.1103/PhysRevA.97.052314} {\bibfield  {journal} {\bibinfo
  {journal} {Phys. Rev. A}\ }\textbf {\bibinfo {volume} {97}},\ \bibinfo
  {pages} {052314} (\bibinfo {year} {2018})}\BibitemShut {NoStop}%
\bibitem [{\citenamefont {Haegeman}\ \emph {et~al.}(2018)\citenamefont
  {Haegeman}, \citenamefont {Swingle}, \citenamefont {Walter}, \citenamefont
  {Cotler}, \citenamefont {Evenbly},\ and\ \citenamefont
  {Scholz}}]{Haegeman2018Rigorous}%
  \BibitemOpen
  \bibfield  {author} {\bibinfo {author} {\bibfnamefont {J.}~\bibnamefont
  {Haegeman}}, \bibinfo {author} {\bibfnamefont {B.}~\bibnamefont {Swingle}},
  \bibinfo {author} {\bibfnamefont {M.}~\bibnamefont {Walter}}, \bibinfo
  {author} {\bibfnamefont {J.}~\bibnamefont {Cotler}}, \bibinfo {author}
  {\bibfnamefont {G.}~\bibnamefont {Evenbly}},\ and\ \bibinfo {author}
  {\bibfnamefont {V.~B.}\ \bibnamefont {Scholz}},\ }\bibfield  {title}
  {\bibinfo {title} {Rigorous free-fermion entanglement renormalization from
  wavelet theory},\ }\href {https://doi.org/10.1103/PhysRevX.8.011003}
  {\bibfield  {journal} {\bibinfo  {journal} {Phys. Rev. X}\ }\textbf {\bibinfo
  {volume} {8}},\ \bibinfo {pages} {011003} (\bibinfo {year}
  {2018})}\BibitemShut {NoStop}%
\bibitem [{\citenamefont {Brouder}\ \emph {et~al.}(2007)\citenamefont
  {Brouder}, \citenamefont {Panati}, \citenamefont {Calandra}, \citenamefont
  {Mourougane},\ and\ \citenamefont {Marzari}}]{Brouder2007Exponential}%
  \BibitemOpen
  \bibfield  {author} {\bibinfo {author} {\bibfnamefont {C.}~\bibnamefont
  {Brouder}}, \bibinfo {author} {\bibfnamefont {G.}~\bibnamefont {Panati}},
  \bibinfo {author} {\bibfnamefont {M.}~\bibnamefont {Calandra}}, \bibinfo
  {author} {\bibfnamefont {C.}~\bibnamefont {Mourougane}},\ and\ \bibinfo
  {author} {\bibfnamefont {N.}~\bibnamefont {Marzari}},\ }\bibfield  {title}
  {\bibinfo {title} {Exponential localization of wannier functions in
  insulators},\ }\href {https://doi.org/10.1103/PhysRevLett.98.046402}
  {\bibfield  {journal} {\bibinfo  {journal} {Phys. Rev. Lett.}\ }\textbf
  {\bibinfo {volume} {98}},\ \bibinfo {pages} {046402} (\bibinfo {year}
  {2007})}\BibitemShut {NoStop}%
\bibitem [{\citenamefont {Monaco}\ \emph {et~al.}(2018)\citenamefont {Monaco},
  \citenamefont {Panati}, \citenamefont {Pisante},\ and\ \citenamefont
  {Teufel}}]{Monaco_2018}%
  \BibitemOpen
  \bibfield  {author} {\bibinfo {author} {\bibfnamefont {D.}~\bibnamefont
  {Monaco}}, \bibinfo {author} {\bibfnamefont {G.}~\bibnamefont {Panati}},
  \bibinfo {author} {\bibfnamefont {A.}~\bibnamefont {Pisante}},\ and\ \bibinfo
  {author} {\bibfnamefont {S.}~\bibnamefont {Teufel}},\ }\bibfield  {title}
  {\bibinfo {title} {Optimal decay of wannier functions in chern and quantum
  hall insulators},\ }\href {https://doi.org/10.1007/s00220-017-3067-7}
  {\bibfield  {journal} {\bibinfo  {journal} {Communications in Mathematical
  Physics}\ }\textbf {\bibinfo {volume} {359}},\ \bibinfo {pages} {61–100}
  (\bibinfo {year} {2018})}\BibitemShut {NoStop}%
\bibitem [{\citenamefont {Cole}\ and\ \citenamefont
  {Vanderbilt}(2025)}]{Cole2025Reduced}%
  \BibitemOpen
  \bibfield  {author} {\bibinfo {author} {\bibfnamefont {T.}~\bibnamefont
  {Cole}}\ and\ \bibinfo {author} {\bibfnamefont {D.}~\bibnamefont
  {Vanderbilt}},\ }\bibfield  {title} {\bibinfo {title} {Reduced wannier
  representation for topological bands},\ }\href
  {https://doi.org/10.1103/PhysRevB.111.205139} {\bibfield  {journal} {\bibinfo
   {journal} {Phys. Rev. B}\ }\textbf {\bibinfo {volume} {111}},\ \bibinfo
  {pages} {205139} (\bibinfo {year} {2025})}\BibitemShut {NoStop}%
\bibitem [{\citenamefont {Qi}\ \emph {et~al.}(2006)\citenamefont {Qi},
  \citenamefont {Wu},\ and\ \citenamefont {Zhang}}]{Qi2006Topological}%
  \BibitemOpen
  \bibfield  {author} {\bibinfo {author} {\bibfnamefont {X.-L.}\ \bibnamefont
  {Qi}}, \bibinfo {author} {\bibfnamefont {Y.-S.}\ \bibnamefont {Wu}},\ and\
  \bibinfo {author} {\bibfnamefont {S.-C.}\ \bibnamefont {Zhang}},\ }\bibfield
  {title} {\bibinfo {title} {Topological quantization of the spin hall effect
  in two-dimensional paramagnetic semiconductors},\ }\href
  {https://doi.org/10.1103/PhysRevB.74.085308} {\bibfield  {journal} {\bibinfo
  {journal} {Phys. Rev. B}\ }\textbf {\bibinfo {volume} {74}},\ \bibinfo
  {pages} {085308} (\bibinfo {year} {2006})}\BibitemShut {NoStop}%
\bibitem [{\citenamefont {Haldane}(1988)}]{Haldane1988Model}%
  \BibitemOpen
  \bibfield  {author} {\bibinfo {author} {\bibfnamefont {F.~D.~M.}\
  \bibnamefont {Haldane}},\ }\bibfield  {title} {\bibinfo {title} {Model for a
  quantum hall effect without landau levels: Condensed-matter realization of
  the "parity anomaly"},\ }\href {https://doi.org/10.1103/PhysRevLett.61.2015}
  {\bibfield  {journal} {\bibinfo  {journal} {Phys. Rev. Lett.}\ }\textbf
  {\bibinfo {volume} {61}},\ \bibinfo {pages} {2015} (\bibinfo {year}
  {1988})}\BibitemShut {NoStop}%
\bibitem [{\citenamefont {Chu}\ \emph {et~al.}(2023)\citenamefont {Chu},
  \citenamefont {Zhu},\ and\ \citenamefont
  {Gorshkov}}]{chu2023entanglementrenormalizationcircuitschiral}%
  \BibitemOpen
  \bibfield  {author} {\bibinfo {author} {\bibfnamefont {S.-K.}\ \bibnamefont
  {Chu}}, \bibinfo {author} {\bibfnamefont {G.}~\bibnamefont {Zhu}},\ and\
  \bibinfo {author} {\bibfnamefont {A.~V.}\ \bibnamefont {Gorshkov}},\ }\href
  {https://arxiv.org/abs/2304.13748} {\bibinfo {title} {Entanglement
  renormalization circuits for chiral topological order}} (\bibinfo {year}
  {2023}),\ \Eprint {https://arxiv.org/abs/2304.13748} {arXiv:2304.13748
  [quant-ph]} \BibitemShut {NoStop}%
\bibitem [{\citenamefont {Chen}\ \emph {et~al.}(2010)\citenamefont {Chen},
  \citenamefont {Gu},\ and\ \citenamefont {Wen}}]{Chen2010Local}%
  \BibitemOpen
  \bibfield  {author} {\bibinfo {author} {\bibfnamefont {X.}~\bibnamefont
  {Chen}}, \bibinfo {author} {\bibfnamefont {Z.-C.}\ \bibnamefont {Gu}},\ and\
  \bibinfo {author} {\bibfnamefont {X.-G.}\ \bibnamefont {Wen}},\ }\bibfield
  {title} {\bibinfo {title} {Local unitary transformation, long-range quantum
  entanglement, wave function renormalization, and topological order},\ }\href
  {https://doi.org/10.1103/PhysRevB.82.155138} {\bibfield  {journal} {\bibinfo
  {journal} {Phys. Rev. B}\ }\textbf {\bibinfo {volume} {82}},\ \bibinfo
  {pages} {155138} (\bibinfo {year} {2010})}\BibitemShut {NoStop}%
\bibitem [{\citenamefont {Hastings}\ and\ \citenamefont
  {Wen}(2005)}]{Hastings2005Quasiadiabatic}%
  \BibitemOpen
  \bibfield  {author} {\bibinfo {author} {\bibfnamefont {M.~B.}\ \bibnamefont
  {Hastings}}\ and\ \bibinfo {author} {\bibfnamefont {X.-G.}\ \bibnamefont
  {Wen}},\ }\bibfield  {title} {\bibinfo {title} {Quasiadiabatic continuation
  of quantum states: The stability of topological ground-state degeneracy and
  emergent gauge invariance},\ }\href
  {https://doi.org/10.1103/PhysRevB.72.045141} {\bibfield  {journal} {\bibinfo
  {journal} {Phys. Rev. B}\ }\textbf {\bibinfo {volume} {72}},\ \bibinfo
  {pages} {045141} (\bibinfo {year} {2005})}\BibitemShut {NoStop}%
\bibitem [{\citenamefont {Bravyi}\ \emph {et~al.}(2010)\citenamefont {Bravyi},
  \citenamefont {Hastings},\ and\ \citenamefont {Michalakis}}]{Bravyi_2010}%
  \BibitemOpen
  \bibfield  {author} {\bibinfo {author} {\bibfnamefont {S.}~\bibnamefont
  {Bravyi}}, \bibinfo {author} {\bibfnamefont {M.~B.}\ \bibnamefont
  {Hastings}},\ and\ \bibinfo {author} {\bibfnamefont {S.}~\bibnamefont
  {Michalakis}},\ }\bibfield  {title} {\bibinfo {title} {Topological quantum
  order: Stability under local perturbations},\ }\bibfield  {journal} {\bibinfo
   {journal} {Journal of Mathematical Physics}\ }\textbf {\bibinfo {volume}
  {51}},\ \href {https://doi.org/10.1063/1.3490195} {10.1063/1.3490195}
  (\bibinfo {year} {2010})\BibitemShut {NoStop}%
\bibitem [{\citenamefont {L{\"o}wdin}(1950)}]{lowdin1950non}%
  \BibitemOpen
  \bibfield  {author} {\bibinfo {author} {\bibfnamefont {P.-O.}\ \bibnamefont
  {L{\"o}wdin}},\ }\bibfield  {title} {\bibinfo {title} {On the
  non-orthogonality problem connected with the use of atomic wave functions in
  the theory of molecules and crystals},\ }\href@noop {} {\bibfield  {journal}
  {\bibinfo  {journal} {The Journal of Chemical Physics}\ }\textbf {\bibinfo
  {volume} {18}},\ \bibinfo {pages} {365} (\bibinfo {year} {1950})}\BibitemShut
  {NoStop}%
\bibitem [{\citenamefont {Des~Cloizeaux}(1964)}]{des1964energy}%
  \BibitemOpen
  \bibfield  {author} {\bibinfo {author} {\bibfnamefont {J.}~\bibnamefont
  {Des~Cloizeaux}},\ }\bibfield  {title} {\bibinfo {title} {Energy bands and
  projection operators in a crystal: Analytic and asymptotic properties},\
  }\href@noop {} {\bibfield  {journal} {\bibinfo  {journal} {Physical Review}\
  }\textbf {\bibinfo {volume} {135}},\ \bibinfo {pages} {A685} (\bibinfo {year}
  {1964})}\BibitemShut {NoStop}%
\bibitem [{\citenamefont {Marzari}\ \emph {et~al.}(2012)\citenamefont
  {Marzari}, \citenamefont {Mostofi}, \citenamefont {Yates}, \citenamefont
  {Souza},\ and\ \citenamefont {Vanderbilt}}]{Marzari2012Maximally}%
  \BibitemOpen
  \bibfield  {author} {\bibinfo {author} {\bibfnamefont {N.}~\bibnamefont
  {Marzari}}, \bibinfo {author} {\bibfnamefont {A.~A.}\ \bibnamefont
  {Mostofi}}, \bibinfo {author} {\bibfnamefont {J.~R.}\ \bibnamefont {Yates}},
  \bibinfo {author} {\bibfnamefont {I.}~\bibnamefont {Souza}},\ and\ \bibinfo
  {author} {\bibfnamefont {D.}~\bibnamefont {Vanderbilt}},\ }\bibfield  {title}
  {\bibinfo {title} {Maximally localized wannier functions: Theory and
  applications},\ }\href {https://doi.org/10.1103/RevModPhys.84.1419}
  {\bibfield  {journal} {\bibinfo  {journal} {Rev. Mod. Phys.}\ }\textbf
  {\bibinfo {volume} {84}},\ \bibinfo {pages} {1419} (\bibinfo {year}
  {2012})}\BibitemShut {NoStop}%
\bibitem [{\citenamefont {Wang}\ \emph {et~al.}(2006)\citenamefont {Wang},
  \citenamefont {Yates}, \citenamefont {Souza},\ and\ \citenamefont
  {Vanderbilt}}]{Wang2006Abinitio}%
  \BibitemOpen
  \bibfield  {author} {\bibinfo {author} {\bibfnamefont {X.}~\bibnamefont
  {Wang}}, \bibinfo {author} {\bibfnamefont {J.~R.}\ \bibnamefont {Yates}},
  \bibinfo {author} {\bibfnamefont {I.}~\bibnamefont {Souza}},\ and\ \bibinfo
  {author} {\bibfnamefont {D.}~\bibnamefont {Vanderbilt}},\ }\bibfield  {title}
  {\bibinfo {title} {Ab initio calculation of the anomalous hall conductivity
  by wannier interpolation},\ }\href
  {https://doi.org/10.1103/PhysRevB.74.195118} {\bibfield  {journal} {\bibinfo
  {journal} {Phys. Rev. B}\ }\textbf {\bibinfo {volume} {74}},\ \bibinfo
  {pages} {195118} (\bibinfo {year} {2006})}\BibitemShut {NoStop}%
\bibitem [{\citenamefont {Kato}(1950)}]{kato1950adiabatic}%
  \BibitemOpen
  \bibfield  {author} {\bibinfo {author} {\bibfnamefont {T.}~\bibnamefont
  {Kato}},\ }\bibfield  {title} {\bibinfo {title} {On the adiabatic theorem of
  quantum mechanics},\ }\href@noop {} {\bibfield  {journal} {\bibinfo
  {journal} {Journal of the Physical Society of Japan}\ }\textbf {\bibinfo
  {volume} {5}},\ \bibinfo {pages} {435} (\bibinfo {year} {1950})}\BibitemShut
  {NoStop}%
\bibitem [{\citenamefont {Bravyi}\ \emph {et~al.}(2011)\citenamefont {Bravyi},
  \citenamefont {DiVincenzo},\ and\ \citenamefont {Loss}}]{Bravyi_2011}%
  \BibitemOpen
  \bibfield  {author} {\bibinfo {author} {\bibfnamefont {S.}~\bibnamefont
  {Bravyi}}, \bibinfo {author} {\bibfnamefont {D.~P.}\ \bibnamefont
  {DiVincenzo}},\ and\ \bibinfo {author} {\bibfnamefont {D.}~\bibnamefont
  {Loss}},\ }\bibfield  {title} {\bibinfo {title} {Schrieffer–wolff
  transformation for quantum many-body systems},\ }\href
  {https://doi.org/10.1016/j.aop.2011.06.004} {\bibfield  {journal} {\bibinfo
  {journal} {Annals of Physics}\ }\textbf {\bibinfo {volume} {326}},\ \bibinfo
  {pages} {2793–2826} (\bibinfo {year} {2011})}\BibitemShut {NoStop}%
\bibitem [{\citenamefont {Hackl}\ and\ \citenamefont
  {Myers}(2018)}]{Hackl_2018}%
  \BibitemOpen
  \bibfield  {author} {\bibinfo {author} {\bibfnamefont {L.}~\bibnamefont
  {Hackl}}\ and\ \bibinfo {author} {\bibfnamefont {R.~C.}\ \bibnamefont
  {Myers}},\ }\bibfield  {title} {\bibinfo {title} {Circuit complexity for free
  fermions},\ }\bibfield  {journal} {\bibinfo  {journal} {Journal of High
  Energy Physics}\ }\textbf {\bibinfo {volume} {2018}},\ \href
  {https://doi.org/10.1007/jhep07(2018)139} {10.1007/jhep07(2018)139} (\bibinfo
  {year} {2018})\BibitemShut {NoStop}%
\bibitem [{\citenamefont {Xie}\ \emph {et~al.}(2026)\citenamefont {Xie},
  \citenamefont {Ono},\ and\ \citenamefont {Po}}]{paperC}%
  \BibitemOpen
  \bibfield  {author} {\bibinfo {author} {\bibfnamefont {W.}~\bibnamefont
  {Xie}}, \bibinfo {author} {\bibfnamefont {S.}~\bibnamefont {Ono}},\ and\
  \bibinfo {author} {\bibfnamefont {H.~C.}\ \bibnamefont {Po}},\ }\bibfield
  {title} {\bibinfo {title} {Fermion gaussian unitary network}} (\bibinfo
  {year} {2026}),\ \bibinfo {note} {manuscript in preparation}\BibitemShut
  {NoStop}%
\bibitem [{\citenamefont {Marzari}\ and\ \citenamefont
  {Vanderbilt}(1997)}]{Marzari1997Maximally}%
  \BibitemOpen
  \bibfield  {author} {\bibinfo {author} {\bibfnamefont {N.}~\bibnamefont
  {Marzari}}\ and\ \bibinfo {author} {\bibfnamefont {D.}~\bibnamefont
  {Vanderbilt}},\ }\bibfield  {title} {\bibinfo {title} {Maximally localized
  generalized wannier functions for composite energy bands},\ }\href
  {https://doi.org/10.1103/PhysRevB.56.12847} {\bibfield  {journal} {\bibinfo
  {journal} {Phys. Rev. B}\ }\textbf {\bibinfo {volume} {56}},\ \bibinfo
  {pages} {12847} (\bibinfo {year} {1997})}\BibitemShut {NoStop}%
\bibitem [{\citenamefont {Cirac}\ and\ \citenamefont
  {Trivedi}(2026)}]{cirac2026efficientrepresentationspepscritical}%
  \BibitemOpen
  \bibfield  {author} {\bibinfo {author} {\bibfnamefont {I.}~\bibnamefont
  {Cirac}}\ and\ \bibinfo {author} {\bibfnamefont {R.}~\bibnamefont
  {Trivedi}},\ }\href {https://arxiv.org/abs/2611.17936} {\bibinfo {title}
  {Efficient representations by peps: critical free-fermion and chiral
  topological states}} (\bibinfo {year} {2026}),\ \Eprint
  {https://arxiv.org/abs/2611.17936} {arXiv:2611.17936 [quant-ph]} \BibitemShut
  {NoStop}%
\bibitem [{\citenamefont {Lieb}\ and\ \citenamefont
  {Robinson}(1972)}]{lieb1972finite}%
  \BibitemOpen
  \bibfield  {author} {\bibinfo {author} {\bibfnamefont {E.~H.}\ \bibnamefont
  {Lieb}}\ and\ \bibinfo {author} {\bibfnamefont {D.~W.}\ \bibnamefont
  {Robinson}},\ }\bibfield  {title} {\bibinfo {title} {The finite group
  velocity of quantum spin systems},\ }\href@noop {} {\bibfield  {journal}
  {\bibinfo  {journal} {Communications in mathematical physics}\ }\textbf
  {\bibinfo {volume} {28}},\ \bibinfo {pages} {251} (\bibinfo {year}
  {1972})}\BibitemShut {NoStop}%
\bibitem [{\citenamefont {Ranard}\ \emph {et~al.}(2022)\citenamefont {Ranard},
  \citenamefont {Walter},\ and\ \citenamefont {Witteveen}}]{Ranard_2022}%
  \BibitemOpen
  \bibfield  {author} {\bibinfo {author} {\bibfnamefont {D.}~\bibnamefont
  {Ranard}}, \bibinfo {author} {\bibfnamefont {M.}~\bibnamefont {Walter}},\
  and\ \bibinfo {author} {\bibfnamefont {F.}~\bibnamefont {Witteveen}},\
  }\bibfield  {title} {\bibinfo {title} {A converse to lieb–robinson bounds
  in one dimension using index theory},\ }\href
  {https://doi.org/10.1007/s00023-022-01193-x} {\bibfield  {journal} {\bibinfo
  {journal} {Annales Henri Poincaré}\ }\textbf {\bibinfo {volume} {23}},\
  \bibinfo {pages} {3905–3979} (\bibinfo {year} {2022})}\BibitemShut
  {NoStop}%
\end{thebibliography}%

\end{document}